\documentclass{aa}  

\usepackage[varg]{txfonts} 
\usepackage{graphicx} 
\usepackage{natbib} 
\bibpunct{(}{)}{;}{a}{}{,} 
 
\usepackage{color}

\begin{document} 
 
   \title{GS242-03+37: a lucky survivor in the galactic gravitational field} 
 
   \author{S. Ehlerov\'a 
          \and 
          J. Palou\v{s} 
          } 
 
   \institute{Astronomical Institute, Academy of Sciences, 
              Bo\v{c}n\'{\i} II 1401, Prague, Czech Republic\\ 
              \email{ehlerova@asu.cas.cz} 
             } 
 
   \date{Received February 28, 2018; accepted August 6, 2018} 
 
   \abstract 
   {HI shells and supershells, found in discs of many galaxies including our own,
    are formed by the activity of young and massive stars (supernova explosions
    and stellar winds), but the formation of these structures may be linked to other energetic events,
    such as interactions of high-velocity clouds with the galactic  disc.
    The larger structures in particular significantly influence their surroundings;
    their walls are often places where molecular clouds reside and where star formation happens.} 
   {We explore the HI supershell GS242-03+37, a large structure in the outer Milky Way.
     Its size and position make it a good case for studying the effects of large shells
   on their surrounding.} 
   {We perform numerical simulations of the structure with the simplified hydrodynamical
     code RING, which uses the thin-shell approximation.
    The best fit is found by a comparison with the HI data and then we compare our model with the
    distribution of star clusters near this supershell.} 
   {The best model of GS242-03+37 requires, contrary to previous estimates,
     a relatively low amount of energy, and it has an old age of $\sim$ 100 Myr. We also find that
     the distribution of young star clusters (with ages $<$ 120 Myr) is correlated with walls of the supershell, while
     the distribution of older clusters is not. Clusters that have the highest probability of being born
     in the wall of the supershell show an age sequence along the wall.}
   {GS242-03+37 is a relatively old structure, shaped by the differential rotation, and its
     wall is a birthplace of several star clusters. The star formation
     started at a time when the supershell was not already supersonically expanding;
       it was a result of the density increase due to the galactic shear and oscillations
       perpendicular to the disc of the Milky Way.}
 
   \maketitle 
 
\section{Introduction}

The interstellar medium (ISM) is full of structures on all scales, from sub-parsec to kiloparsec. 
In the Milky Way, HI shells ranging in sizes from a few parsec to about 1 kpc were discovered first by  
\citet{heiles1979} and later by others:
\citet{mcclure2002}, \citet{2005A&A...437..101E}, \citet{2013A&A...550A..23E}, and \citet{suad2014}. 
HI shells were also found in other galaxies \citep[for a review on shells in external galaxies 
and an analysis of HI shells in THINGS galaxies, see][]{bagetakos+2011}. 
 
Since the time of their discovery it has been speculated that HI shells are the result of the energetic activities 
of massive stars: winds, radiation, and supernova (SN) explosions. For many large shells, 
the energy involved in their creation must have come from the whole cluster of stars; from
an OB association. Proving  the connection between shells and clusters is not completely
straightforward since many shells are older than the expected lifetime of massive stars and
the results of such comparisons are often confusing and even contradictory
\citep{1999AJ....118..323R,2000ApJ...529..201S}.
 The shells are like footprints remaining in the ISM after the cluster is gone.
They should be able to survive encounters with random density fluctuations and
stay coherent when they are deformed by the sheer due to galactic differential rotation. 
The statistical correspondence between  HI shells and the CO distribution 
\citep{ehlerova2016} indicates
that shells may trigger the formation of molecular clouds and new stars.

Many observational papers, both on the Milky Way and on external galaxies, 
show that HI shells frequently exist at large galactocentric distances, far from star-forming regions and often
with relatively large sizes. This implies that alternative mechanisms 
might be employed to explain these structures: ram pressure \citep{2002AJ....123.1316B} 
or the infall of a high-velocity cloud to the disc \citep{1987A&A...179..219T}. 
These events might be responsible for a fraction of HI shells, but probably not 
for the majority. 
 
HI shells evolve in gaseous discs. Once their dimensions are  comparable to the disc thickness
--- which is not unusual --- their shapes should be influenced by the large-scale density gradient  
in the disc and they may prolong in the z-direction (these prolonged shells are usually called 
worms). If the interior of the shell is still hot, that is, if the massive progenitor stars still exist, 
this hot gas may flow into the galactic halo (the worms become chimneys).  
In such a way HI shells may influence energetic flows in galaxies. 
For dwarf galaxies with low gravity \citep[e.g.][]{2009A&A...493..511V} 
or for starburst galaxies (e.g. M82), such an event could mean the loss of 
the hot gas to the intergalactic medium.

A number of HI shells fall into the category of supershells. Supershells are larger
than `normal' shells, but above all they are more energetic, that is the energy
needed to create them is much larger than one (or a few) supernova explosions.
Sometimes the energy requirements for supershells are as high as 1000 SNe. That is why supershells
are prime candidates for having a non-stellar origin, since only
exceptionally large star-forming regions are able to produce such a large amount
of energy.

Walls of shells and supershells are denser than their surroundings and are therefore places
where star formation is likely to be triggered. \citet{dawson2011, dawson2013}
and \citet{ehlerova2016} studied the connection between the (super)shells and
the level of molecularization in the Large Magellanic Cloud (LMC) and in the outer Milky Way
and found that supershells slightly increase the amount of molecules (and therefore
slightly increase the star formation) compared to the situation without the shells. A recent
example of the star formation triggered in the HI shell in the Magellanic Bridge is given
by \citet{mackey2017}. Strange arcs of clusters in the LMC (Sextant and Quadrant)
are also sometimes considered to be created by the HI shell
\citep[perhaps by LMC-4, as suggested in][]{efremov1999},
but other theories have also been invoked
\citep[for a review concerning the origin of Sextant and Quadrant, see][]{efremov2013}.
Physical processes and timescales involved in the triggered star formation, including the
star formation triggered in HI shells, can be found in \citet{elmegreen2011}.

GS242-03+37 is an HI supershell observed in the outer Milky Way. It is a dominant object
observed in the HI maps of the Galaxy \citep[see][for a detailed description of this structure]{mcclure2006}.
As one of the still rather enigmatic supershells,
and as one of striking HI features, it merits being studied to better estimate its properties
and to analyse its influence on its surroundings, both gaseous and stellar.

The paper is organised as follows: 
section \ref{sec:data} describes datasets used in the paper, section \ref{sec:gs242}
deals with previous observations of the supershell GS242-03+37 and its properties.
Section \ref{sec:ringcode} gives an overview of the numerical code used for calculations.
Section \ref{sec:models} compares numerical models with observations of GS242-03+37.
Section \ref{sec:clusters} analyses the distribution of clusters in the vicinity of GS242-03+37 and
section \ref{sec:summary} provides a summary.

\section{Data}
\label{sec:data}

In our study we use HI and CO radio observations and further information about the
star clusters in the area.

\begin{figure}
\centering
\includegraphics[angle=-90,width=0.9\linewidth]{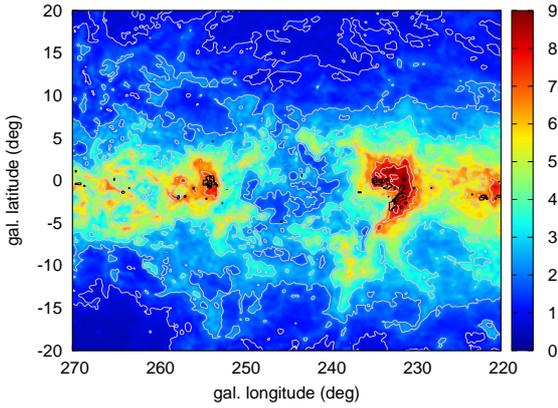}
\caption{Galactic supershell GS242-03+37: channel map integrated between
  +30 and +50 $\mathrm{kms^{-1}}$ (HI4PI survey). Thick black lines
  are the CO emission, which is only available for the strip $b \in (-5^{\circ},+5^{\circ})$.
  \citep{2001ApJ...547..792D}. 
     }
       \label{gs242-lb}
\end{figure}

\begin{figure*}
\centering
\includegraphics[angle=-90,width=0.45\linewidth]{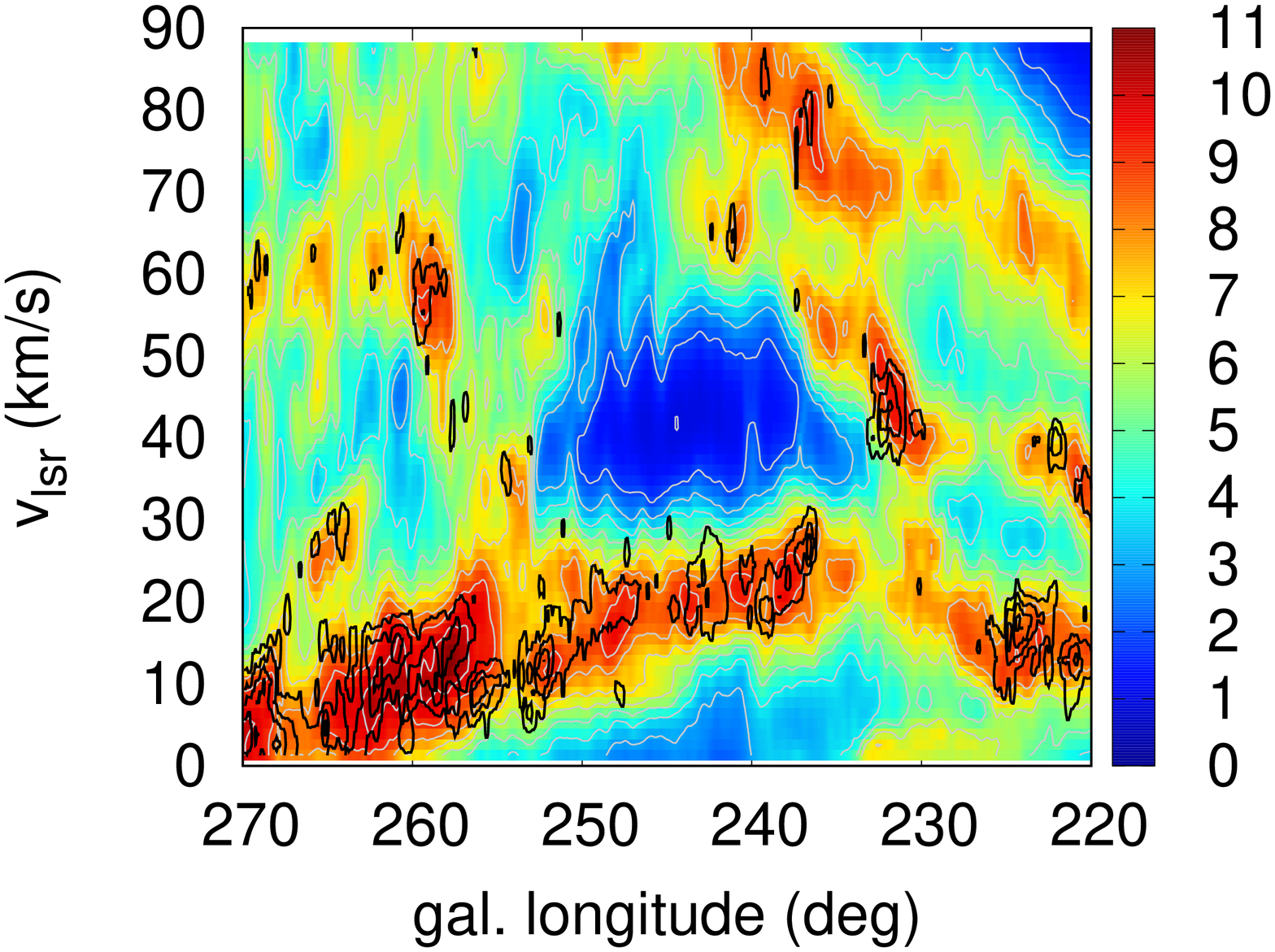}
\includegraphics[angle=-90,width=0.45\linewidth]{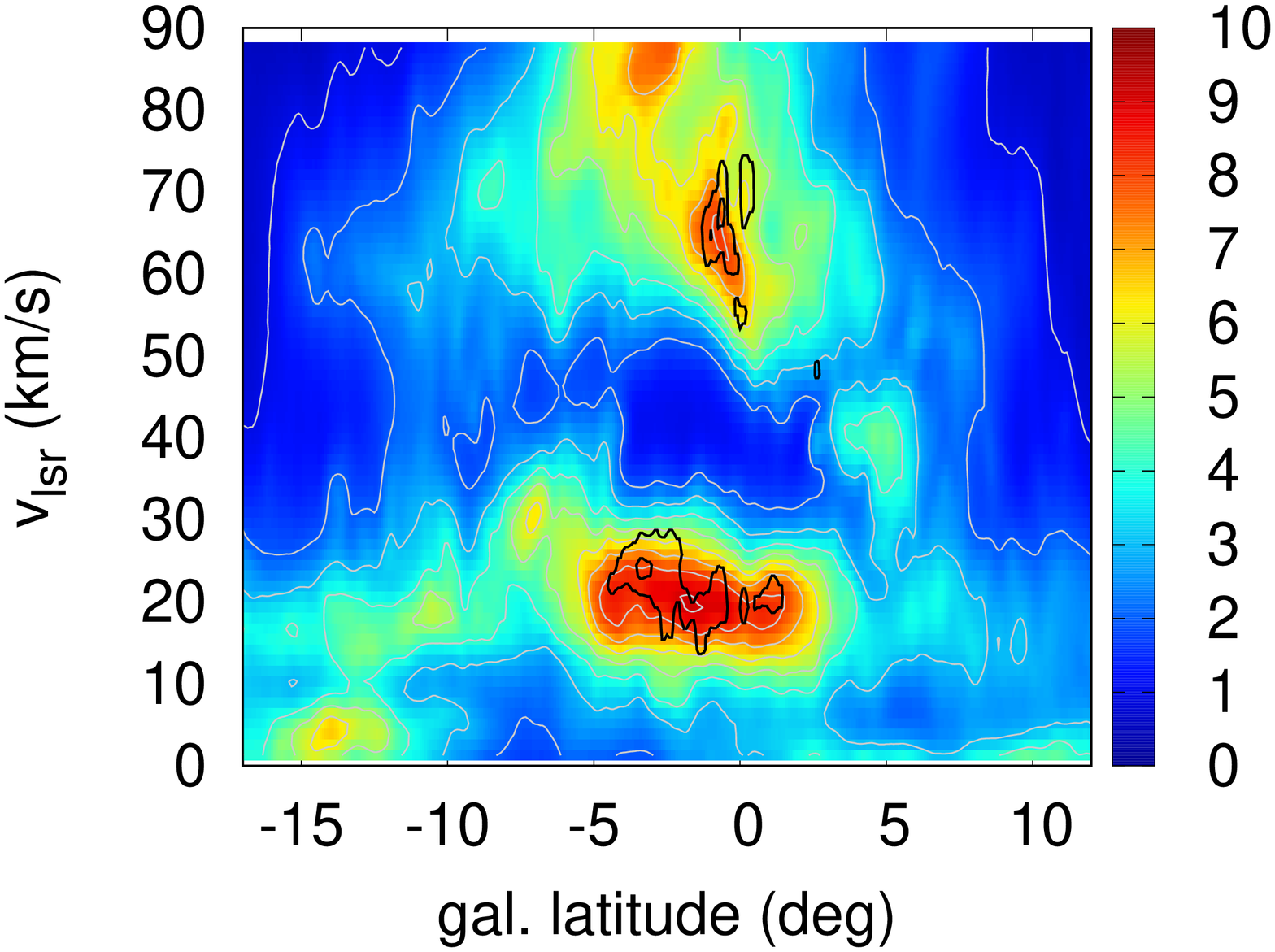}
\caption{PV-diagrams of the shell GS242-03+37: the longitude--velocity diagram
  integrated between $b = -1^{\circ}$ and $-3^{\circ}$ (left) and
  the latitude--velocity integrated between $l = 241^{\circ}$ and  $245^{\circ}$ (right).
  Thick black lines have the same significance as in Fig.1.
     }
       \label{gs242-lvbv}
\end{figure*}

\subsection{HI and CO}

We use the all-sky HI4PI survey \citep{hi4pi2016}. This survey is based on the Effelsberg-Bonn
HI Survey (EBHIS) for the northern hemisphere and the Galactic All-Sky Survey (GASS) for the
southern hemisphere. The angular resolution of the HI4PI is $\theta_{\mathrm{FWHM}} = 16^{\prime}.2$,
the spectral resolution is $1.29\ \mathrm{kms}^{-1}$, the sensitivity is $\sigma_{\mathrm{rms}} \simeq 43 \ mK$.

For CO we use the CO (J = 1-0) survey of \citet[][]{2001ApJ...547..792D}. It is
a composite survey of the Milky Way, consisting of observations from several telescopes.
In our studied region it covers the strip around the Galactic equator between $b \in (-5^{\circ}, +5^{\circ})$.
The combined datacube  (the so-called deep CO survey) that we use has a pixel 
size of $0.125^{\circ}$ and a channel width of $\Delta v = 1.3\ \mathrm{kms^{-1}}$,
the root mean square (rms) noise is 0.1 K.

\subsection{Clusters}

We take positions, distances, and ages of clusters from the catalogue of \citet{kharchenko+2013}.
The advantage of this catalogue -- and the reason why we use it -- is its homogeneity.  
Properties of clusters in this catalogue are calculated from the
pipeline, which starts with the positions of clusters from previous
catalogues, takes the stellar properties of stars from stellar catalogues,
calculates the membership probability of individual stars,
and using the theoretical stellar tracks derives the basic properties of the cluster.
The catalogue is estimated to be complete to the heliocentric distance 1.8 kpc. 

In our paper we use coordinates of clusters, their heliocentric distances, ages and
(if available) the radial velocities\footnote{In the whole paper, when using the radial velocity,
we mean the velocity relative to the local standard of rest, $v_{\mathrm{lsr}}$.}.
\citet{kharchenko+2013} state that the relative error of the distance determination is 11 \%, and
that the relative error of the age determination
is between 25 \% (internal error, i.e. an error resulting from using the pipeline itself) and 39 \%
(external error, i.e. from the comparison with other sources, which includes the errors of these
other sources).

\section{GS242-03+37} 
\label{sec:gs242}

GS242-03+37 is a large prominent object, which dominates the HI distribution in the third
galactic quadrant. It is visible in the velocity range $\sim (+20,+65)\ \mathrm{kms^{-1}}$ between
$l \in (230^{\circ}, 255^{\circ})$ and $b \in (-7^{\circ}, +5^{\circ})$ (see Figs. \ref{gs242-lb}
and \ref{gs242-lvbv}).
We start with the description of the HI gas around the supershell, then focus on the
HI image of the supershell and its properties and  describe the distribution
of star clusters in this area. We end this section with remarks about other observed Galactic supershells.

\begin{figure*}
\centering
\includegraphics[angle=-90,width=0.45\linewidth]{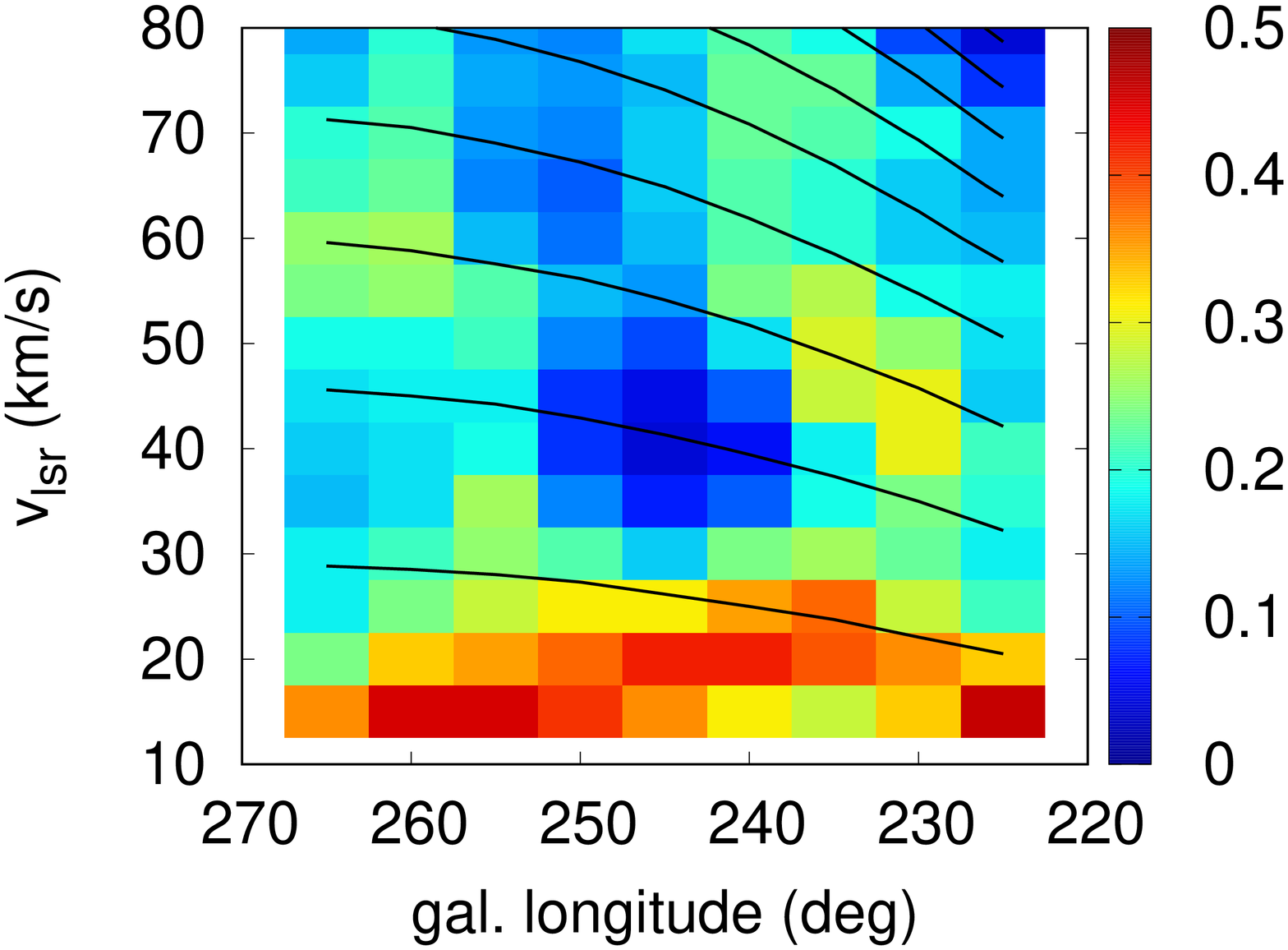}
\includegraphics[angle=-90,width=0.45\linewidth]{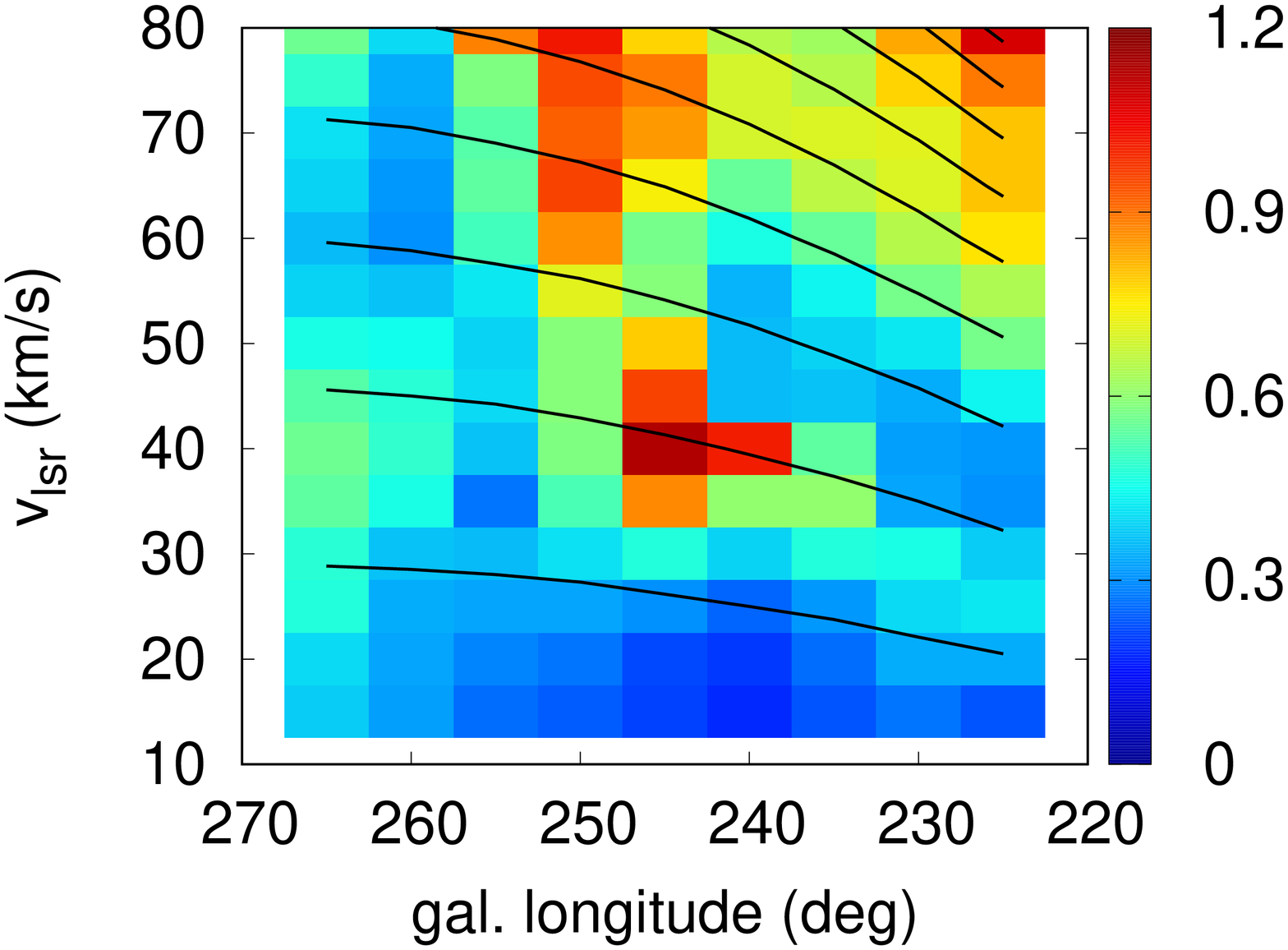}
\caption{
  Maximum density of the HI gas (left panel) and its thickness (right panel) based on simple
  conversion from HI data to the Gaussian HI gas distribution using the rotation curve.
  Black lines show galactocentric distances, starting from 10 kpc (the lowest line)
  with the 1 kpc increment. Densities are in $at\ cm^{-3}$, thicknesses are in kpc.
  The supershell GS242-03+37 is clearly visible on both maps.
     }
       \label{HIdensmap}
\end{figure*}

\subsection{HI distribution around GS242-03+37}

There should be three spiral arms in the direction of $l \simeq 240^{\circ}$: the Local arm,
the Perseus arm and the Outer arm. Unfortunately, while their presence is something most
authors agree on, their distances are not that well known. Based on \citet{koo2017} and
\citet{reid2014} we estimate the distance to the Local arm as being between 1 and 2 kpc, the
distance to the Perseus arm as around 5 kpc and the Outer arm around 8 kpc (our assumed
galactocentric distance of the Sun is $R_{\odot} = 8.5$ kpc). 
The most probable radial velocities connected with these arms are +20 $\mathrm{kms^{-1}}$ for
the Local arm, +65 $\mathrm{kms^{-1}}$ for the Perseus arm, and +90 $\mathrm{kms^{-1}}$ for the
Outer arm; but the ranges are wide, especially for the Perseus arm, which is bifurcated
in the area \citep{koo2017}. 

The supershell GS242-03+37 at $v_{LSR} \simeq 37\ kms^{-1}$ therefore lies in the inter-arm region.
The blueward (i.e. approaching)
wall coincides with --- or lies in --- the Local arm (or the HI overdensity which
is associated with the Local arm). The redward (receding) wall might or might not coincide with
the Perseus arm: the situation is complex, as obvious from Fig. \ref{gs242-lvbv}, because the
exact location of the spiral arm is not clear, but it is probably located at slightly
larger radial velocities than the receding wall of our structure. The clear image,
signs of the expansion in $lb$ maps, and the presence of walls and a significant hole in $lv$ and $bv$
diagrams all strengthen the claim that the GS242-03+37 is a real structure, not just an inter-arm region.

We can estimate the HI density distribution from the HI brightness distribution
by assuming the rotation curve (or more generally, the velocity field).
We warn that this is only a very rough estimate, since 1) we deal with the expanding structure
which involves non-rotational velocity \textit{per se}, and 2) all other disturbances (e.g. spiral arms)
add non-rotational velocities and therefore induce errors. Nevertheless, we can try and take the results,
but with great caution. We use the rotation curve of \citet{wouterloot1990} 
\begin{equation} 
  V (R) = V_{\odot} \left({R \over R_{\odot}}\right)^{0.0382}, 
  \label{rotcurve} 
\end{equation}   
where $R$ and $V$ are the galactocentric distance and rotation velocity. The subscript $_{\odot}$ denotes
the values at the position of the Sun, which we assume to be $R_{\odot} = 8.5$ kpc and
$V_{\odot} = 220\ kms^{-1}$.

We do the conversion from temperatures $T_{\mathrm{B}}$ to HI densities
by summing {and averaging} the HI emission in cubes of  $5^{\circ}$ in
$l$ and $5\ \mathrm{kms}^{-1}$ in $v$
and by fitting the Gaussian distribution of the HI density to average $T_{\mathrm{B}}$ values
for all these cubes for $b$ between $-30^{\circ}$ and $+30^{\circ}$.
Results of the fitting are the maximum density $n_0$, the position of the maximum density
$z_0$ and the thickness of the HI layer $\sigma_{HI}$ for each position $l$ and $v$.
Maximum HI density $n_0$ and the Gaussian thickness $\sigma_{HI}$ for different positions
in the Galaxy are plotted in Fig. \ref{HIdensmap}, together with distances derived using the rotation curve given by Eq. \ref{rotcurve}.

Figure \ref{HIdensmap} shows the expected global decrease of density with the galactocentric
distance $R_{\mathrm{GC}}$ and the increase of the HI layer thickness with the increasing $R_{\mathrm{GC}}$.
The warp (i.e. the offset between the nominal Galactic equator and a position
of the maximum density in the HI layer) is also growing with the galactocentric distance
(not shown).
At the position of GS242-03+37 we see a hole in the density $n_0$, which is at least partially
real. Non-rotational velocities, which are not accounted for, may also contribute to
  the decrease of the density at the position of the expanding structure.
A second interesting result
is the local increase of the disc thickness at the position $l = 242^{\circ},
  v_{LSR} = 37\ \mathrm{kms}^{-1}$, with the maximum density slightly below the plane $b=0^{\circ}$.
The disc appears inflated;
 this is in correspondence with the assumption of an energetic event which significantly
disturbed the disc and which has not yet had time to relax.

\subsection{Galactic HI supershell GS242-03+37}

\begin{figure}[b]
\centering
\includegraphics[angle=-90,width=0.9\linewidth]{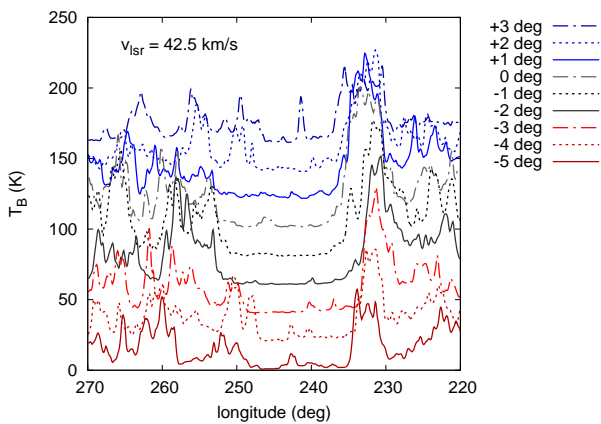}
\caption{Cuts of the HI channel map at $v_{\mathrm{lsr}} = 42.5 \mathrm{kms}^{-1}$ along the lines
  of constant galactic latitude $b$. Individual profiles are artificially offset by 20 K.}
       \label{tbprofile}
\end{figure}

The galactic shell GS242-03+37 was first discovered and described by \citet{heiles1979},
who also classified it as a supershell. It was later reidentified (in different data sets) by
other authors, for example by \citet{2005A&A...437..101E}, \citet{2013A&A...550A..23E},
and \citet{mcclure2006}, but sometimes also
not found, as in \citet{suad2014}; though they do identify many smaller structures in the vicinity
of the supershell.
The structure is visible in the velocity range $\sim (+20,+65)\ \mathrm{kms^{-1}}$ between
$l \in (230^{\circ}, 255^{\circ})$ and $b \in (-7^{\circ}, +5^{\circ})$ (see Figs. \ref{gs242-lb}
and \ref{gs242-lvbv}). The lower boundary in $b$ is slightly fuzzy and there are several
possible walls; our best tip is around $-7^{\circ}$.
The densest parts of the HI supershell contain the CO emission.

\citet{mcclure2006} describe the supershell GS242-03+37 as it looks in HI observations
and other wavelengths. The main difference between their description and ours is the angular
dimension of the structure, both in $l$ and $b$. The angular dimension of the supershell
in \citet{mcclure2006} is $15^{\circ}$, while ours is $25^{\circ}$. The smaller dimension is
based, as is evident from our Figs. \ref{gs242-lb} and \ref{gs242-lvbv} or from Fig. 3 in
\citet{mcclure2006}, on the size of the hole, that is, the empty or low-density region.

Our estimate of the supershell size is based on the distribution of high-density walls and arcs around
the low-density hole. Figure \ref{tbprofile} shows cuts of one HI channel map along the lines of constant
galactic latitude $b$. $T_{\mathrm{b}}$ profiles share many similarities (i.e. the low-brightness temperature
  region surrounded by higher-temperature walls), but there are many individualities in profiles. Some profiles
show large temperature contrasts between the wall and the hole and the high steepness of the profile
in the part corresponding to the wall (e.g. for $b = 0^{\circ}$ and the wall around $l \simeq 230^{\circ}$),
but others are less steep and have lower contrast. The wall at $l \simeq 260^{\circ}$ is particularly
fragmented, which is certainly connected to the ongoing star formation in this region, as we demonstrate below.
Walls shown in Fig. \ref{tbprofile} are relatively thick. The $v_{\mathrm{lsr}}$ velocity of the shown channel
(42.5 $\mathrm{kms}^{-1}$) was chosen as one of those channels, where the
supershell has the largest angular dimension. Nevertheless, profiles for another channel would show
similar properties \citep[e.g.][gives another example of a high contrast $T_{\mathrm{B}}$ profile]{mcclure2006}.
We define the angular size in $l$ as the maximum of the angular distances between the maxima
of the brightness temperature $T_B$ in the opposite walls in cuts across the shell,
which yields a value of $25^{\circ}$.    

The second difference between \citet{mcclure2006} and our image of GS242-03+37 is the openness or elongation
of the structure in the z direction. \citet{mcclure2006} find HI caps at latitudes of $\sim +20^{\circ}$
and between $\sim -10^{\circ}$ and $\sim -20^{\circ}$ (a part of this cap is marginally visible in
Fig. \ref{gs242-lb}). They conclude that GS243-03+37 is partially open to the galactic halo and that it forms
several chimneys which channel the hot gas to high $z$.
We cannot exclude the connection of observed higher $b$ arcs to GS242-03+37, but
we put a greater weight on the existence of (broken) arcs at latitudes of
$\sim +5^{\circ}$,  $\sim -7^{\circ}$ and  $\sim -10^{\circ}$. Because these walls exist, the
structure cannot be a fully blown-out shell. However, it is possible that the high-latitude
caps were created by the hot gas leaking from the main body of the supershell, which nevertheless
managed  to avoid being completely broken during this process.

The usually adopted heliocentric distance of GS242-03+37 is 3.6 kpc, which corresponds to a radial
velocity of $38\ \mathrm{kms^{-1}}$ for our rotation curve. For this distance, the diameter of the shell
is 1.7 kpc for the angular size of $25^{\circ}$ (or 1 kpc for $15^{\circ}$)
and the FWHM thickness of walls is (120-250) pc.
Estimating the shell radius as half of its angular dimension and expansion velocity as half
of its velocity extent we get $r_{\mathrm{sh}} = 850\ \mathrm{pc}$ and $v_{\mathrm{exp}} = 20\ \mathrm{kms}^{-1}$.
The volume density of the ambient gas is $n_{0} \simeq 0.2\ \mathrm{cm}^{-3}$ (see Fig. \ref{HIdensmap}).

The energy that created the shell can be estimated either from the measured kinetic
energy (and assuming the ratio of the total to kinetic energy), or from the so-called
Chevalier's formula \citep[Eq. \ref{chevalier};][]{chevalier1974}, which
is based on MHD evolution of spherically symmetric supernova remnants, unfortunately
without many physical processes. Therefore it is probably not accurate, but is a possible starting point.
\begin{equation}
  \left( E_{\mathrm{tot}} \over \mathrm{erg} \right ) =
  5.3 \times 10^{43} \left( n_0 \over \mathrm{cm^{-3}} \right )^{1.12}
  \left( v_{\mathrm{exp}} \over \mathrm{kms^{-1}} \right )^{1.40}
  \left( r_{\mathrm{sh}} \over \mathrm{pc} \right )^{3.12}
  \label{chevalier}
,\end{equation}
where $E_{\mathrm{tot}}$ is the total energy involved in the creation of the structure,
$n_0$ is the density of the ambient medium, $r_{\mathrm{sh}}$ the radius of the shell,
and $v_{\mathrm{exp}}$ its expansion velocity.

Both these methods give an estimate of  $\sim 10^{54}\ \mathrm{erg}$ (or thousands
of supernovae) as the energy needed to create the supershell GS242-03+37.
However, as rightly noted by  \citet{mcclure2006}, the shell probably has the lower expansion
velocity ($v_{exp} = 7\ kms^{-1}$ rather than $20\ kms^{-1}$); a part of the velocity extent of the shell (visible in
Fig. \ref{gs242-lvbv}) is caused by the galactic rotation, which substantially decreases the energy
$E_{\mathrm{tot}}$ required to create the supershell.

An interesting property of the shell is its age. From analytical solutions
(see the Section \ref{sec:ringcode}) it follows that
$t = 0.4 r_{\mathrm{sh}}/ v_{\mathrm{exp}}$ (Sedov solution) or
$t = 0.6 r_{\mathrm{sh}}/ v_{\mathrm{exp}}$ \citep{weaver1977}.
Therefore, taking approximately one half of the ratio between the radius of the observed
structure and its expansion velocity seems to be a reasonable estimate of its age,
which is sometimes called the dynamical age of the shell.
\begin{equation}
  t_{\mathrm{dyn}} = 0.5 {r_{\mathrm{sh}} \over v_{\mathrm{exp}}}
  \label{dyntime}
,\end{equation}
where $t_{\mathrm{dyn}}$ is in Myr, $r_{\mathrm{sh}}$ in pc and $v_{\mathrm{exp}}$ in $\mathrm{kms}^{-1}$.
For GS242-03+37, this gives $\sim $20 Myr (or 60 Myr for $v_{exp} = 7\ kms^{-1}$).

\begin{figure}
\centering
\includegraphics[angle=-90,width=0.9\linewidth]{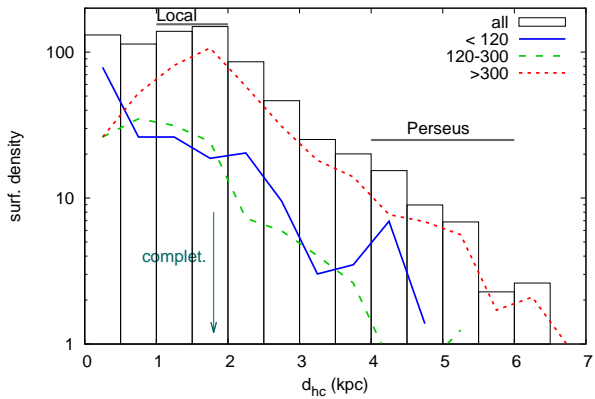}
\caption{Surface density of clusters in the region ($l \in (225^{\circ},260^{\circ})$)
  from \citet{kharchenko+2013} as a function of the heliocentric distance $d_{\mathrm{hc}}$.
  The histogram shows the density of all clusters; lines indicate the distribution of clusters with
  a certain age (blue: clusters younger than 120 Myr, green: clusters between 120 and 300 Myr,
  red: clusters older than 300 Myr). Positions of spiral arms (Local and Perseus) are indicated,
  as well as the limiting distance, where the catalogue is complete.
     }
       \label{clusters-dhc}
\end{figure}
  
  \subsection{Clusters around GS242-03+37}

Figure \ref{clusters-dhc} shows the heliocentric distances of clusters in the region
$l \in (224^{\circ},260^{\circ})$, $b \in (-40^{\circ},+30^{\circ})$. Up to 1.8 kpc the catalogue is
claimed to be complete
and the surface density in this interval is nearly constant, about 120 clusters per $\mathrm{kpc}^{-2}$.
From this distance onwards the density is decreasing, both because of the increasing incompleteness
and of the real decrease in the surface density (the increasing heliocentric distance in this
direction means also increasing galactocentric distance). In Fig. \ref{clusters-dhc} there
are differences in the distribution of clusters based on their ages.
The distribution shows no obvious or unambigious signs of the presence of the spiral arm.

\subsection{Other supershells in the Milky Way}

GS242-03+37 is not the only supershell observed in the Milky Way.
A detailed description of some of the supershells and attempts to identify
the energy source can be found in \citet{maciejewski1996}, \citet{pidopryhora2007}
and \citet{park2016}. The Aquila supershell \citep{maciejewski1996} has an estimated energy of
$10^{52}-10^{53}$ erg deposited by SN explosions and stellar winds. Another example is
the Ophiuchus superbubble \citep{pidopryhora2007} with an estimated total energy input of $\sim 10^{53}$ erg,
which also originates from SNe and winds. The supershell GS040.2+00.6-80 is connected with a high-velocity
cloud and probably created by its infall \citep{park2016}. The first two objects are located in the
inner Milky Way, the third one in the outer Galaxy.

Molecular emission from walls of supershells GSH287+04-17 (the Carina Flare supershell) and
GSH277+00+36 is studied in \citet{dawson2011} and \citet{mcclure2003};
for the Carina Flare supershell also in \citet{2012A&A...539A.116W}.

\section{Expansion of shocks in the ISM}
\label{sec:ringcode}

\begin{figure*}
\centering
\includegraphics[angle=-90,width=0.45\linewidth]{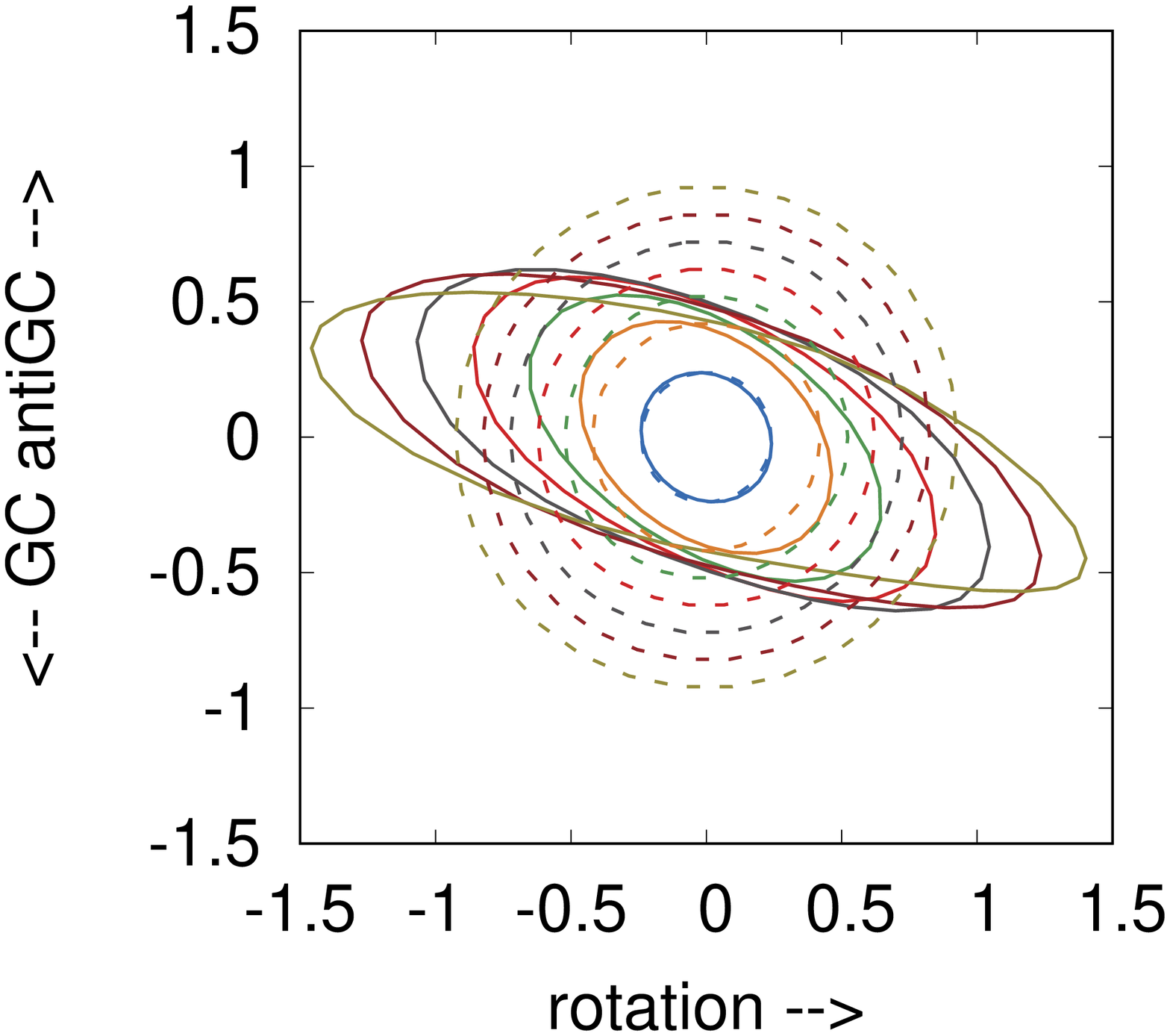}
\includegraphics[angle=-90,width=0.45\linewidth]{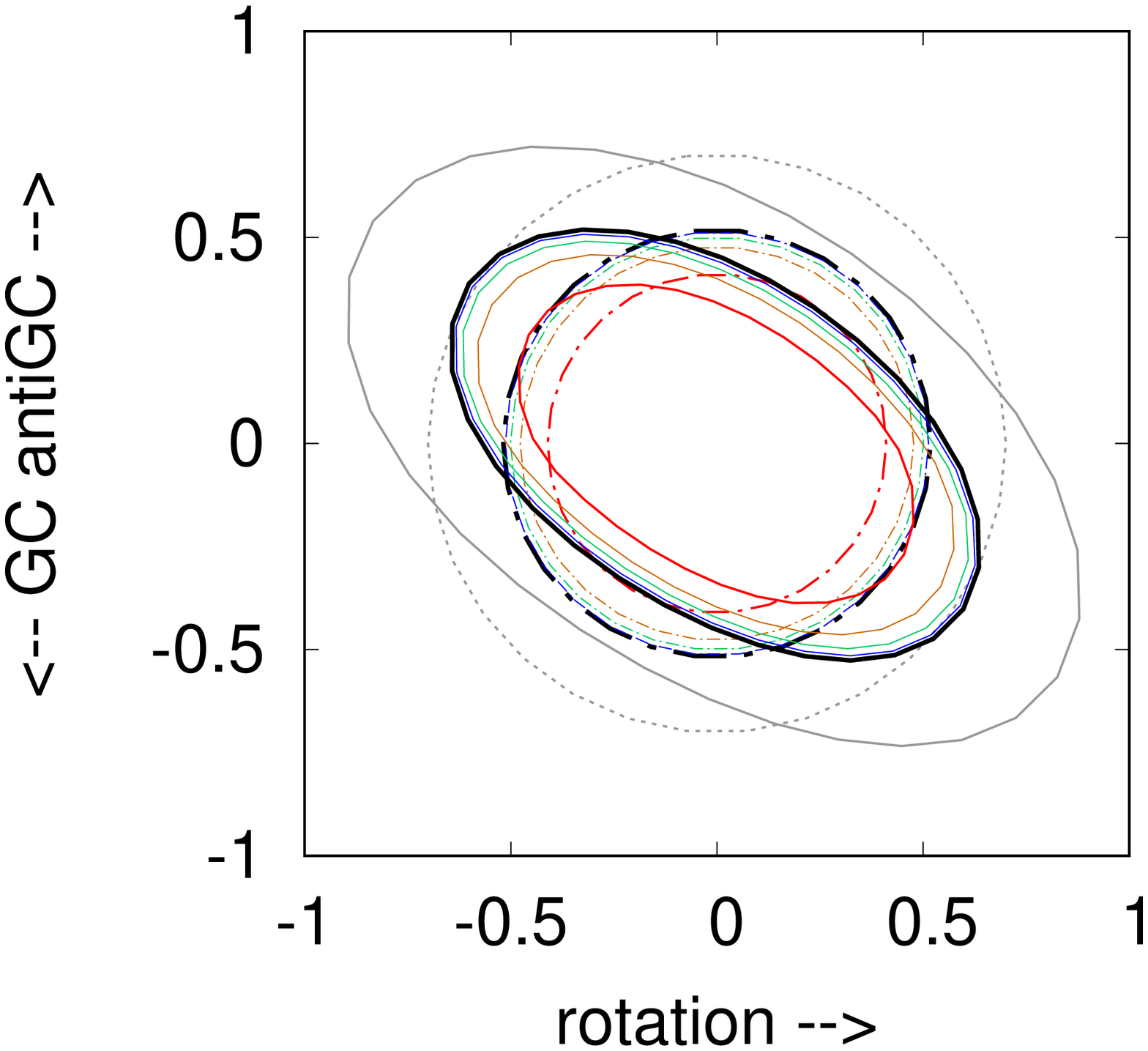}
    \caption{The evolution of the shell calculated by the RING code; cuts are in the galactic
plane (10, 30, 50, 70, 90, 110 and 130 Myr; left), dashed lines correspond to the model
without the galactic potential, solid lines show the model with the potential. The right panel 
shows the evolutionary time 50 Myr
for different models: homogeneous medium (black), homogeneous medium with 5x larger energy input (grey),
Gaussian density disc with scale heights of 500 pc (blue), 300 pc (green), 200 pc (brown) and 100 pc (red).
Dashed and dotted lines 
show the situation without the galactic potential, solid lines the situation with the galactic potential.
Axes are labelled in kpc.
     }
       \label{xy-evolution}
\end{figure*}

\begin{figure*}
\centering
\includegraphics[angle=-90,width=0.45\linewidth]{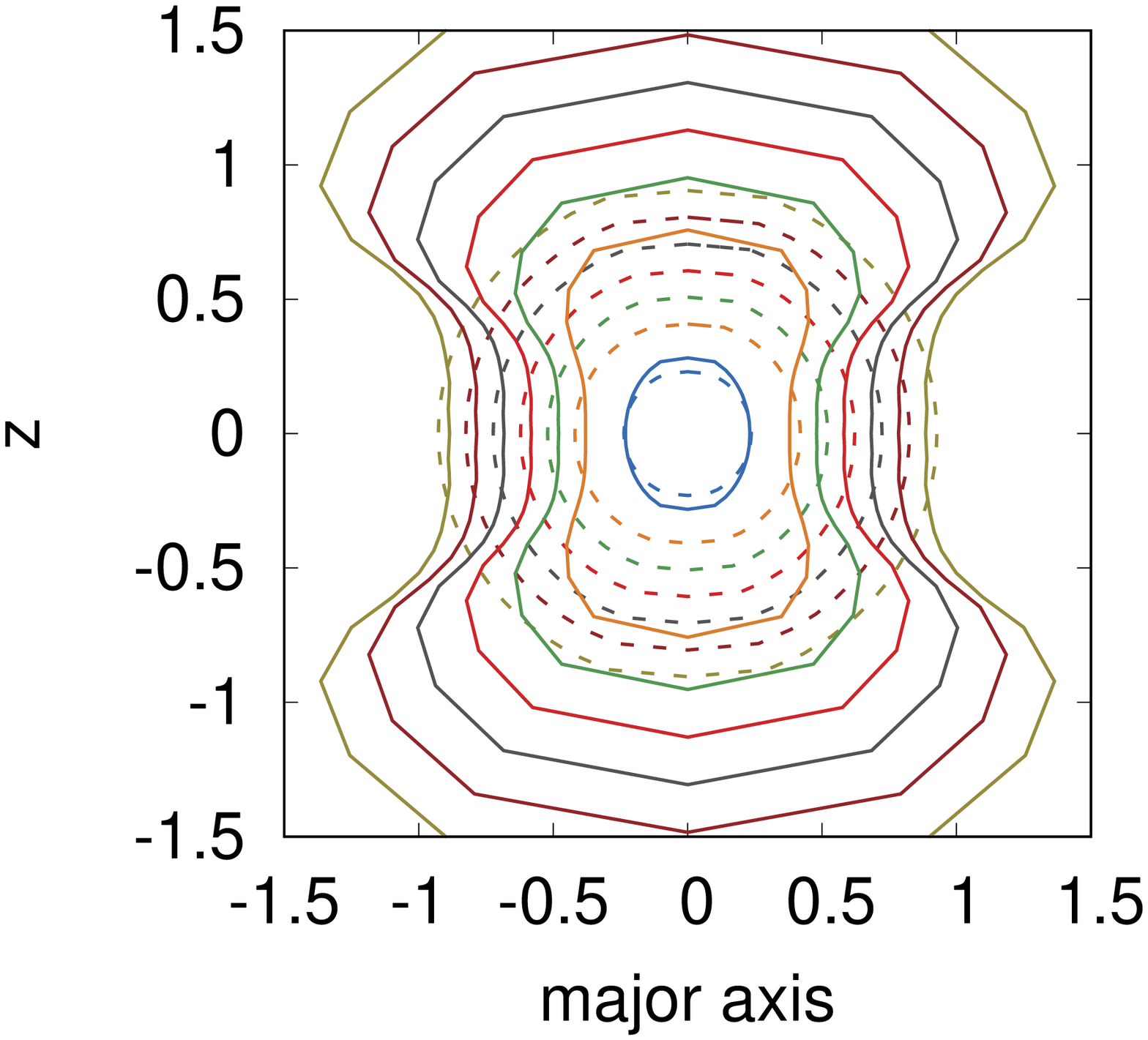}
\includegraphics[angle=-90,width=0.45\linewidth]{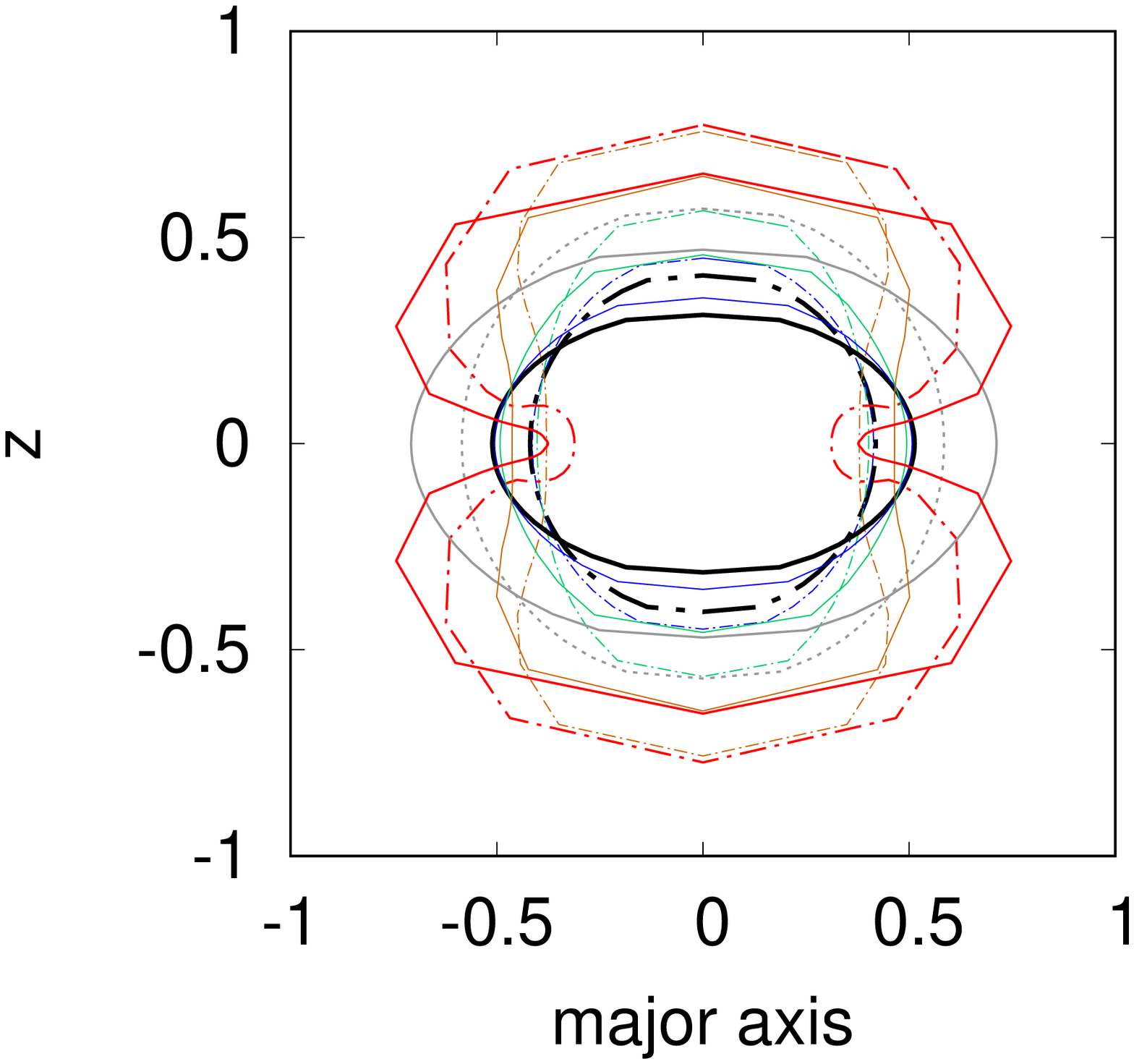}
    \caption{The evolution of the shell calculated by the RING code; cuts are perpendicular to the galactic
plane (10, 30, 50, 70, 90, 110 and 130 Myr; left). Dashed lines corresponds to the evolution in the 
homogeneous medium, solid lines show the Gaussian profile with the scaleheight of 200 pc. No galactic
potential is taken into account. The right panel shows the evolutionary time 30 Myr
for different models: homogeneous medium (black), homogeneous medium with 5x larger energy input (grey), Gaussian density
disc with scale heights of 500 pc (blue), 300 pc (green), 200 pc (brown) and 100 pc (red).
Dashed and dotted lines show the situation without the galactic potential, solid lines the situation
with the galactic potential. Axes are labelled in kpc.
     }
       \label{xz-evolution}
\end{figure*}

\begin{figure}
\centering
\includegraphics[angle=-90,width=0.9\linewidth]{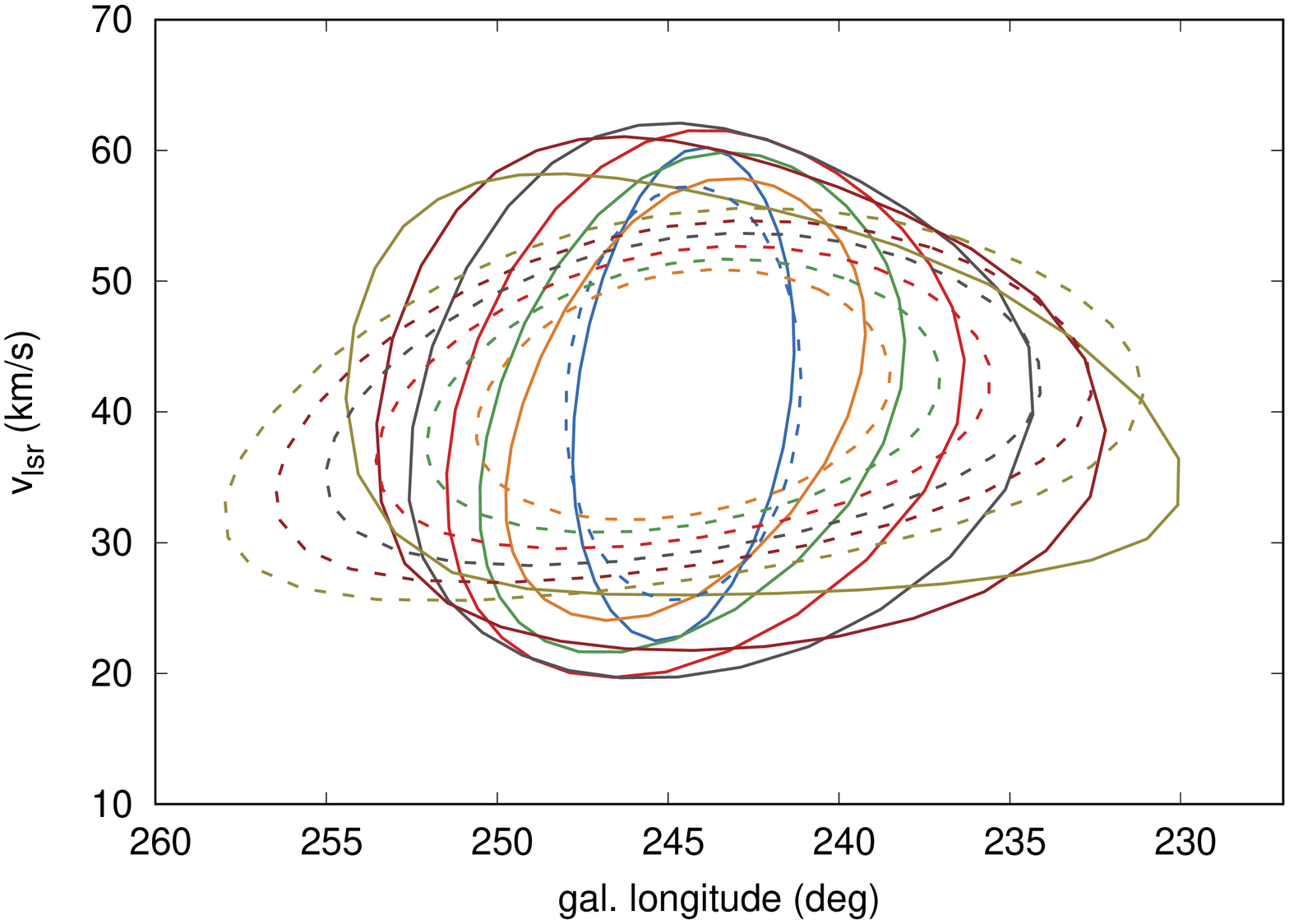}
\caption{The evolution of the shell calculated by the RING code, $lv$-diagram,
  at times (10, 30, 50, 70, 90, 110 and 130) Myr. Dashed lines correspond to the model
  without the galactic potential, solid lines to the model with the potential.
     }
       \label{lv-evolution}
\end{figure}

When referring to the shell in this section and beyond, we refer to the whole structure, consisting
of the (dense) wall and the (low-density) interior. We start with
a general introduction to theoretical models of blast waves, continue with the 
description of our numerical model, and then present some results based on our simulations.

\subsection{General introduction}

The evolution of blast waves, either resulting from supernovae or stellar winds,
expanding in the ISM has been described in many papers and books, for example 
by \citet{ostriker1988} and \citet{bisnovatyj-kogan1995}.
Here we summarise very briefly how the blast wave, resulting from the supernova
explosion, evolves.
The first stage of the supernova explosion is the phase of the free expansion.
 This is followed by the so-called Sedov-Taylor phase characterised by the constant
thermal energy inside the blast wave driving the expansion.
When the radiative losses in the shell of the swept-up gas become important
and the work done on the interstellar gas is thus radiated away, the dense wall of the
shell behind the shock shrinks to a thin layer and the structure
enters another phase. When the pressure inside the shell drops,
the wave is not driven by the interior pressure anymore and it only keeps its
momentum -- this is the so-called snowplough phase. Later, when the expansion velocity
decreases below a local sound speed, the thin wall moves balistically in the gravitational
field of the galaxy.

The evolution of the shell in the Sedov-Taylor phase is described by the
self-similar Sedov solution:
\begin{equation} 
  r_{\mathrm{sh}} =
  \left( {25 \over 4 \pi }\right)^{1 \over 5} \left({E_0 \over \rho_{\mathrm{out}}}\right)^{1 \over 5} t^{2 \over 5} 
\label{sedov-r}
\end{equation}  
\begin{equation} 
  v_{\mathrm{exp}} =
  {2 \over 5} \left( {25 \over 4 \pi }\right)^{1 \over 5} \left({E_0 \over \rho_{\mathrm{out}}}\right)^{1 \over 5} t^{-{3 \over 5}}
\label{sedov-v}
,\end{equation} 
where $r_{\mathrm{sh}}$ is the radius of the shell, $v_{\mathrm{exp}}$ its expansion velocity,
$t$ is the time since the beginning of the expansion, $E_0$ is the energy of the explosion,
and $\rho_{\mathrm{out}} = \mu n_0$ is the  mass density of the ambient medium
($\mu$ is the average mass of the particle).

It is possible to derive solutions for other phases of the blast wave evolution
\citep[we refer to][chapter 8]{2003adu..book.....D}. Here we give the
solution derived by \citet{weaver1977} for models with a continuous energy supply; for example,
for shells powered by stellar winds.

\begin{equation} 
  r_{\mathrm{sh}} =
  \left( {25 \over 14 \pi }\right)^{1 \over 5} \left({L_{\mathrm{source}} \over \rho_{\mathrm{out}}}\right)^{1 \over 5} t^{3 \over 5} 
\label{weaver-r}
\end{equation}

\begin{equation} 
  v_{\mathrm{exp}} =
  {3 \over 5} \left( {25 \over 14 \pi }\right)^{1 \over 5} \left({L_{\mathrm{source}} \over \rho_{\mathrm{out}}}\right)^{1 \over 5} t^{-{2 \over 5}}
\label{weaver-v}
,\end{equation}
where $L_{\mathrm{source}}$ is the energy flux from the source.

In both these sets of equations (Eqs. \ref{sedov-r} and \ref{sedov-v} plus \ref{weaver-r} and \ref{weaver-v}),
the parameter is the ratio between the inserted internal energy $E_{th, in}$ or an energy flux $L_{source}$
and the density
$\rho_{out}$. This degeneracy makes it difficult to estimate both of these quantities from
measurable shell sizes and expansion velocities.

\subsection{Description of the RING code}

Our model assumes that the wall of the shell is infinitesimally thin, that is, its thickness 
is much smaller than its diameter.
This approach was developed by \citet{kompaneets1960},
and \citet{bisnovatyj-kogan1982} and used by \citet{tenorio-tagle1987}, 
\citet{ehlerova1996}, \citet{silich1996} and others.

The time evolution of the wall is calculated in three dimensions. The wall is divided into 
$n_\mathrm{l}$ layers and every layer into $n_\mathrm{p}$ elements (there are typically
around 1000 elements altogether).

The expansion starts from a small spherical volume. The total initial mass 
is the same as what would be necessary to fill the initial volume 
with the density of the surrounding  ISM $\rho_{\mathrm{out}}$ at the centre of explosion,
this mass is distributed evenly to all elements.
Into this small initial volume we insert the initial thermal energy 
$E_{\mathrm{th, in}} = {5 \over 11} E_0$, and the shell gets the initial kinetic energy 
$E_{\mathrm{kin, sh}} = {15 \over 77} E_0$ \citep[the ratios are based on the wind solution by][]{weaver1977}. 
The remaining ${27 \over 77} E_0$ is the initial thermal energy of the shell, which  
(as explained above) is radiated away. 

The thermal energy of the shell interior $E_{\mathrm{th, in}}$ follows the energy balance equation 
\begin{equation} 
  {dE_{\mathrm{th, in}} \over dt} = L_{\mathrm{source}} - {dV_{\mathrm{in}} \over dt} P_{\mathrm{in}}, 
\label{energy} 
\end{equation} 
where $L_{\mathrm{source}}$ is the energy flux from the source (young stars or supernovae) and
$V_{\mathrm{in}}$ and $P_{\mathrm{in}}$ are the volume of the shell interior and its pressure.
Our adopted thin-shell approximation
assumes that the thin shell behind the leading shock has such a high density
that it is able to radiate away all the thermal energy produced by compression of the ambient
medium by the shock (the last term in Eq. \ref{energy}). On the other hand, cooling of 
the low-density medium in the shell interior is very small and is disregarded.

The pressure inside the shell follows the equation: 
\begin{equation} 
P_{\mathrm{in}} = {2 \over 3} {E_{\mathrm{th, in}} \over V_{\mathrm{in}}} 
.\end{equation}
The energy is supplied continuously and constantly with a flux $L_{\mathrm{source}}$  during the time
interval $t_{\mathrm{energy}}$. When the time $t > t_{\mathrm{energy}}$,  the energy supply is stopped.
The time interval $t_{\mathrm{energy}}$ is one of the free parameters in our model. If $t_{\mathrm{energ}y}$
equals zero, all energy $E_{\mathrm{th, in}}$ is delivered abruptly at the beginning
of the expansion.

For each element of the thin wall we solve the momentum and mass equations. The momentum
equation is
\begin{eqnarray} 
  {d \over dt}(m_{\mathrm{i}} \vec{v_{\mathrm{i}}})
  + \rho_{\mathrm{out}} 
  [{\vec{S_{\mathrm{i}}} . (\vec{v_{\mathrm{i}}}} - \vec{v_{\mathrm{out}}})] \vec{v_{\mathrm{out}}}
  = (P_{\mathrm{in}} - P_{\mathrm{out}})\vec{S_{\mathrm{i}}}  + m_{\mathrm{i}}\vec{g}
  \label{momentum} 
,\end{eqnarray}   
where $m_{\mathrm{i}}$ is the mass of an ith element, 
      $\vec{S_{\mathrm{i}}}$ is its surface (i.e. the surface multiplied by the normal vector),  
      $\vec{v_{\mathrm{i}}}$ is its velocity of expansion, 
      $\vec{v_{\mathrm{out}}}$ is the velocity of the medium outside of the shell, 
      $P_{\mathrm{in}}$ and $P_{\mathrm{out}}$ are the inside and outside pressures, 
      $\rho_{\mathrm{out}}$ is the density of the medium outside of the shell, and 
$\vec{g}$ is the gravitational acceleration in the Milky Way.

When the expansion of the shell is supersonic, the supershell collects the ambient medium.
For each element, when the velocity component perpendicular to the shell surface
$v_{\perp} = (\vec{v_{\mathrm{i}}} - \vec{v_{\mathrm{out}}})_{\perp}$ exceeds the local speed of sound,
the mass accumulation is given as 
\begin{equation} 
  {dm_{\mathrm{i}} \over dt} =
  \rho_{\mathrm{out}} \ [(\vec{v_{\mathrm{i}}} - \vec{v_{\mathrm{out}}}) . \vec{S_{\mathrm{i}}}]
  \label{mass} 
.\end{equation}   
When $v_{\perp}$ drops below the local sound speed the accumulation of mass is stopped. From
that time on the element continues its expansion in the gravitational potential of the galaxy
in the ballistic way. Individual elements of the shell are completely independent of each
other; they are not influenced by the low-density hole inside, nor do they influence it.
 
In a homogeneous medium with no external forces from the Milky Way, our numerical solution agrees 
with the solution given by \citet{weaver1977}. 
 
In the Milky Way, shells evolve in the differentially rotating disc, 
which is described by the rotation curve \citep[Eq. \ref{rotcurve},][]{wouterloot1990}.
The rotation curve defines the two components of the gravitational acceleration parallel to the
galactic disc.  The component of the gravitational acceleration perpendicular to it, $g_z$, is
approximated by the formula \citep{kuijken1989} 
\begin{equation} 
  g_z = -2 \pi G z  \left[{\Sigma_D \over (z^2 + z_D^2)^{1/2}} + 2 \rho_H\right], 
  \label{gzpotential} 
\end{equation}       
where $z$ is the distance perpendicular to the galaxy plane, $z_D$ is the disc 
thickness, $\Sigma_D$ is the disc surface density, and $\rho_D$ is the 
halo density. We adopted the following values: 
$z_D = 300$ pc, $\Sigma_D = 46$ M$_{\odot}$ pc$^{-2}$, and 
$\rho_H = 0.015$ M$_{\odot}$pc$^{-3}$. 
 
The distribution of the ISM is a free boundary condition. In this paper we use homogeneous or Gaussian
distribution with the midplane density and the thickness as parameters. More specific distributions,
such as \citet{dickey1990}, can be easily used in the code,
however, for the purposes of this paper the single component disc was sufficient.
We do not take into account the small-scale clumpiness and the chaotic nature of ISM; we assume these
inhomogeneities flatten out on a larger scale.

\subsection{Examples of simulated expanding supershells and the wisdom learned}

\begin{figure}
\centering
\includegraphics[angle=-90,width=0.9\linewidth]{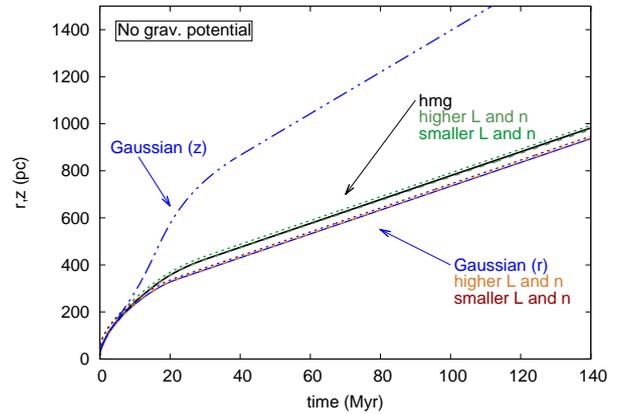}
\caption{The radius $r_{\mathrm{sh}}$ of the shell as a function of time,  calculations of the RING code,
  and evolution without the external gravitational potential.
  Black and green lines correspond to the evolution in the homogeneous medium,
  blue and red lines to the evolution in the Gaussian disc with the thickness
  of 200 pc.
  Black line is the basic model, green lines (dashed and dotted, respectively) correspond to models with
  10x higher and smaller energy inputs with adequately higher and smaller
  densities ($L/n$ are the same for all models). Blue lines show the radius of the
  shell in the plane (solid) and the $z$-extent (dash-dotted) for the basic
  model. Red and orange lines again correspond to models with
  higher and smaller energy inputs and densities, respectively.
     }
       \label{nopot-evolution}
\end{figure}

\begin{figure*}
\centering
\includegraphics[angle=-90,width=0.45\linewidth]{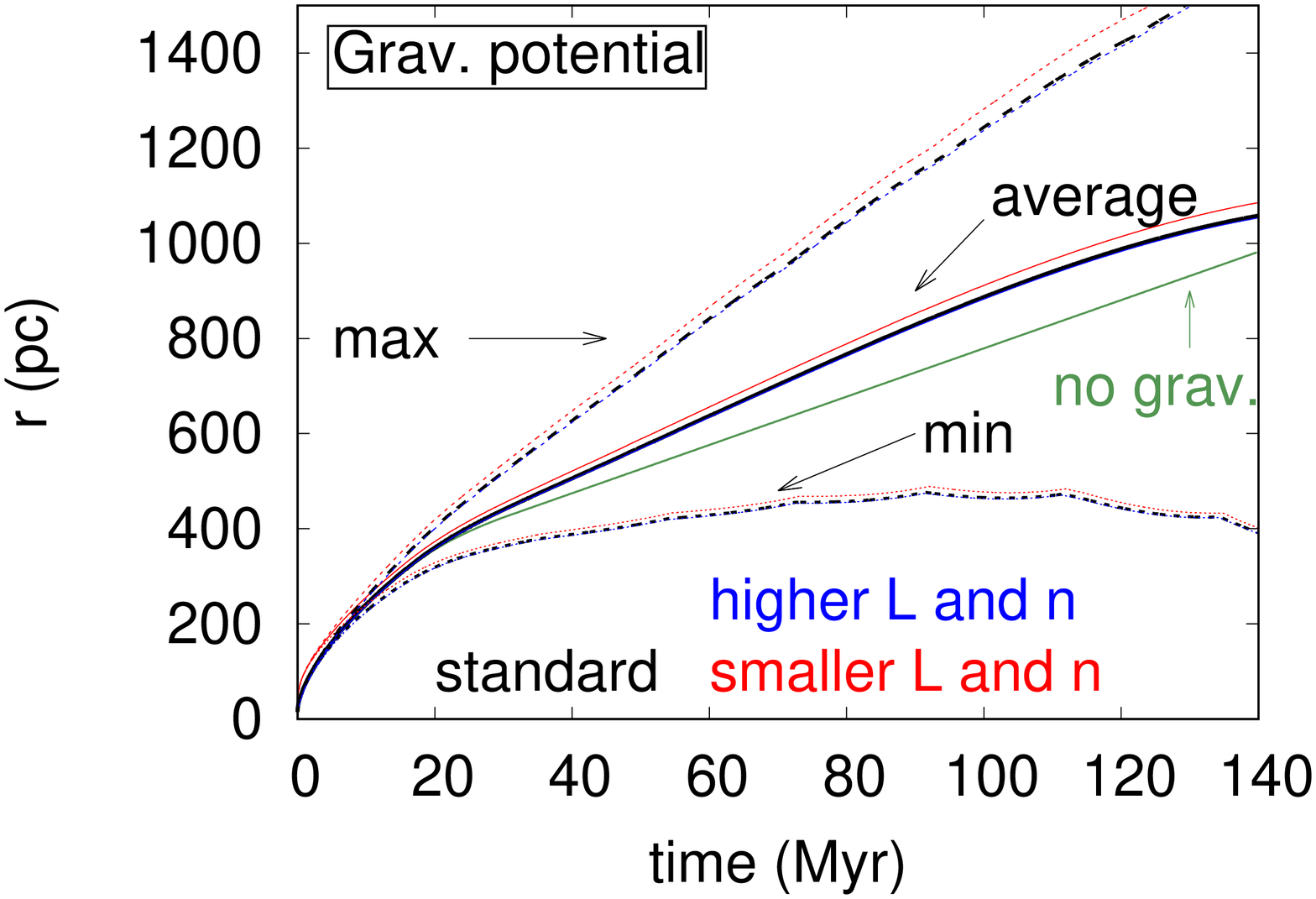}
\includegraphics[angle=-90,width=0.45\linewidth]{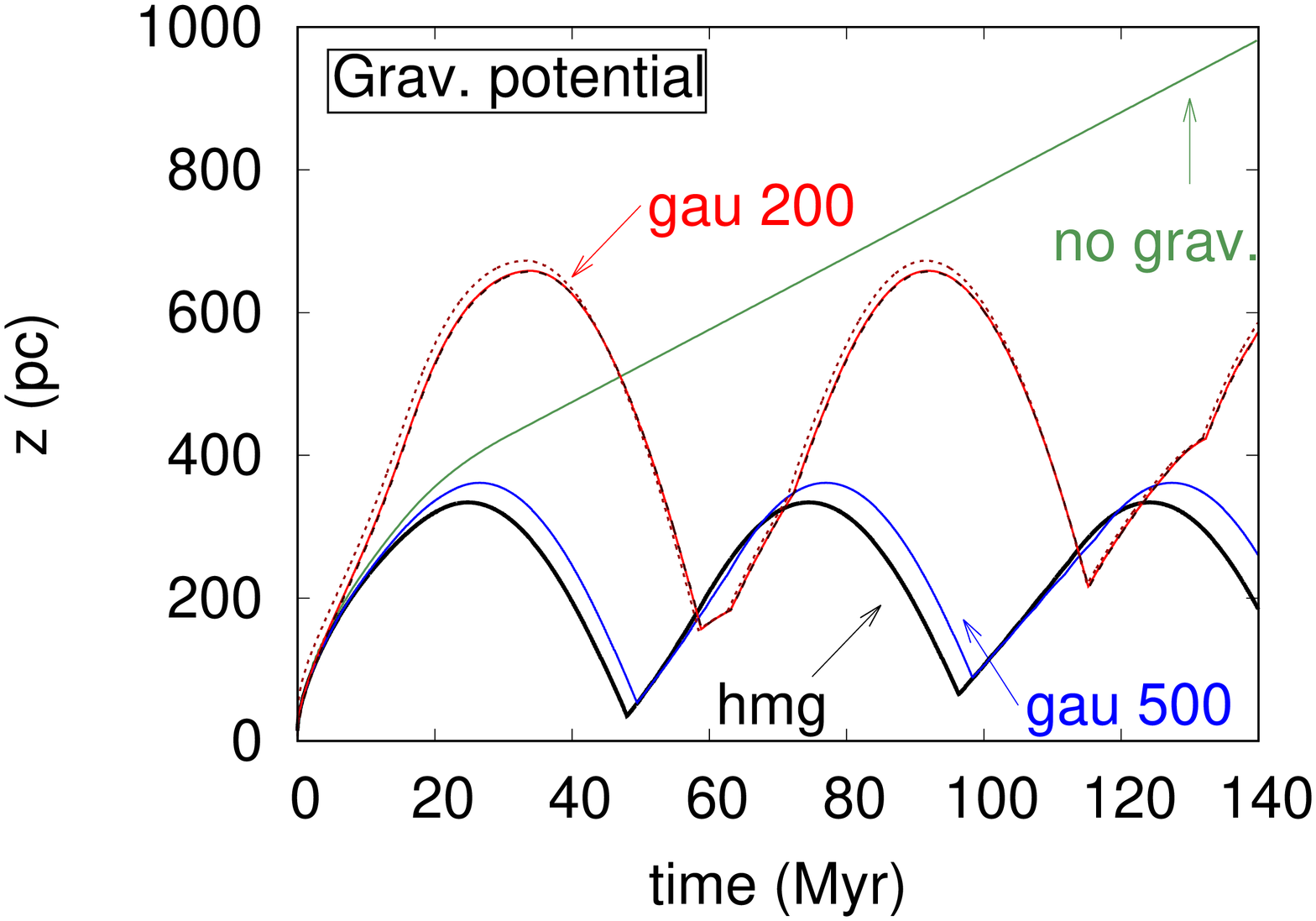}
\caption{Radius $r_{\mathrm{sh}}$ of the shell as a function of time,
  calculations of the RING code, the evolution with the external gravitational potential:
  the radius in the plane (left) and the $z$-extent (right). Left: 
  black colour corresponds to the basic model, blue colour
  to the model with 10x higher energy input and density, red colour to the model
  with 10x lower energy input and density. Different line types are dedicated for the average values
  $ r_{\mathrm{ave}}$ (solid), the maximum $r_{\mathrm{max}}$ (dashed) and minimum $r_{\mathrm{min}}$
  (dotted) values. Right: the $z$-extent
  of the basic model in the homogeneous medium (black), the Gaussian distribution
  with a thickness 500 pc (blue) and the Gaussian distribution with a thickness
  of 200 pc (red; models with 10x higher and lower energy input and density are
  added). In both panels the basic model from Fig. \ref{nopot-evolution}
  is overlaid in green.
     }
       \label{pot-evolution}
\end{figure*}

There are three quantities that dominate the evolution of expanding shells:
1) the density distribution, that is, the stratification of the ISM;
2) the galactic gravitational potential, that is, the rotation velocity
Eq. \ref{rotcurve} and the $g_z$ potential Eq. \ref{gzpotential}; and
3) the amount of energy added to the shell.
In this section we try to present the most important conclusions showing the dependence of
observable quantities on these three input parameters (the density stratification,
the galactic potential, and the energy input).

Models described in this section and shown in figures are calculated for the mid-plane density
$0.2\ cm^{-3}$ and the energy input of $13 \times 10^{51}\ \mathrm{erg}$ delivered
during 15 Myr (these rather strange numbers are chosen because, as shown later,
they are parameters of the best fit to the studied supershell).
The galactic potential -- if used -- corresponds to the galactocentric distance of 11 kpc.

Figures \ref{xy-evolution} and \ref{xz-evolution} show the time evolution of an expanding supershell.
Figure \ref{xy-evolution} gives the cuts through the shell at the galactic plane $z$ = 0,
and Fig. \ref{xz-evolution} shows the shape of the shell in the direction perpendicular to the disc.

Figure \ref{xy-evolution} (left) shows an evolution in the homogeneous medium
with and without the external gravitational potential. While the shell evolving
without the external gravitational field remains spherical, the shape and orientation of the
structure in the external field change due to the differential galactic rotation, the shell
becomes more and more twisted with time. At young ages (10 Myr) this influence
is small, but at later times (> 50 Myr) this distortion, and the difference from the 
case without the galactic differential rotation, becomes significant. 
  
The right panel of Fig. \ref{xy-evolution} shows the outlines of different models
at the evolutionary time 50 Myr: two homogeneous
cases (the one in the left panel and the other one with 5x the energy input) and
shells evolving in Gaussian density discs with scale heights of 500 pc, 300 pc, 200 pc
and 100 pc. For all cases, models with and without the galactic potential are shown.
There is no significant difference between models evolving in the homogeneous medium
and in thick discs (scale heights larger than or equal to 200 pc). Larger energy inputs
create larger shells (as expected). Shells in thin discs are slightly smaller than
the shells with the same energy evolving in the thicker discs. This is due to the
expansion to the galactic halo.

Figure \ref{xz-evolution} shows the influence of the density stratification of the gaseous disc.
The left panel shows the evolution of two models: a shell in the homogeneous density
and a shell evolving in the Gaussian disc with a scale height of 200 pc. No galactic
potential is included. In the case of the evolution in the homogeneous medium the shell 
remains spherical and its expansion slows down. In the case of the stratified disc,
the shell becomes elongated along the density gradient (it becomes a `worm' in the terminology
of HI shells). Its its
upper and lower parts may accelerate and the blowout may occur (in the shown case,
the blowout is not really prominent, but the structure is significantly elongated).
The elongation of the structure caused by the density gradient is significant already at early times.

The right panel of Fig.\ref{xz-evolution} illustrates the effects of the galactic
acceleration $g_z$ at the evolutionary time 30 Myr. Six different models are shown: two homogeneous
cases (the one in the left panel and the other one with 5x the energy input) and
shells evolving in Gaussian density discs with scale heights of 500 pc, 300 pc, 200 pc
and 100 pc. For all cases, models with and without the galactic potential are shown. 
Due to the galactic $g_z$ potential shells are more flattened (compared to no-potential case).
At later times, the individual parts
of the structure begin to oscillate around the galactic plane (see Fig. \ref{pot-evolution}, right)
and the z-profile of the shell ceases to hold any information on the density
stratification it had during the young age.

According to analytical solutions of Sedov (Eqs. \ref{sedov-r} and \ref{sedov-v}) 
and Weaver (Eqs. \ref{weaver-r} and \ref{weaver-v}) describing shells in a
homogeneous medium without the external gravitational field, quantities $r_{\mathrm{sh}}$ and $v_{\mathrm{sh}}$ 
depend on the ratio $E_0/n_{\mathrm{out}}$ or $L_{\mathrm{source}}/n_{\mathrm{out}}$.
Figures \ref{nopot-evolution} and \ref{pot-evolution} study these degeneracies for shells evolving
in stratified density discs and with the external gravitational potential.
The basic model is
shown together with a model with ten times higher energy flux and density and
with a model with ten times lower energy flux and density; all three models  having the same
ratio $L_{\mathrm{source}}/n_{\mathrm{out}}$. Figures show that
the ratio $L_{\mathrm{source}}/n_{\mathrm{out}}$ is really the dominating quantity
and that the degeneracies derived for simple models appear also
in stratified discs with gravitational potential.

Shells evolving in the differentially rotating and z-stratified galactic discs are no longer spherical.
We characterise them with the maximum $r_{\mathrm{max}}$, minimum $r_{\mathrm{min}}$ and average $r_{\mathrm{ave}}$
radii in the $z = 0$ plane, and with the maximum $z$ distance of the shell caps (see Fig. \ref{pot-evolution}
for the time evolution).
The average radius $r_{\mathrm{ave}}$ is calculated as the average value of the radius along the whole
perimeter of the shell at the galaxy plane ($z = 0$).The value of   $r_{\mathrm{ave}}$ computed with the gravitational
potential does not differ significantly from the case without the potential, but $r_{\mathrm{min}}$ and $r_{\mathrm{max}}$
are significantly influenced (Fig. \ref{pot-evolution}, left), since the shell evolving without the external
gravitational potential remains spherical. 

The right panel of Fig. \ref{pot-evolution} shows the time evolution of the maximum $z$-extent
of the shell under
the influence of the gravitational potential of the Galaxy. After the initial stage of
the supersonic shell expansion, the trajectories of shell fragments oscillate around the
Galactic plane. This motion increases the wall density at times when fragments
are close to the plane. Trajectories are synchronised because they start from initial
conditions given by the expanding shell and therefore evolve in a similar way.
The period of oscillations depend on the z-profile of the gravitational potential, which depends on the
surface density of the (stellar) disc and on the halo density
(through Eq. \ref{gzpotential}), and also on parameters of the shell.

An important quantity is the thickness of the gaseous disc.
Shells in a non-stratified disc or shells in thick discs ($\sim 500\ pc$) evolve in a very similar manner.
If the shell is significantly elongated ($\sim 200\ pc$) or even if it ventilates a part of its energy
to the halo
(for shells in thin discs), the evolution is slightly different. It takes a longer
time before the attractive force of the galactic disc overwhelms the expansion
for the highest (or lowest) parts of the shell (= caps) --- in extreme cases these parts
  can be thrown away from the disc and from the galaxy, but these cases are not described in this paper.

These $z$-oscillations do not influence the in-plane evolution of the shell, since they occur
in later times of the shell evolution, when the motion is subsonic and when the shell
is no longer driven by the interior pressure (as in Eq. \ref{momentum}) and when it no longer accumulates
new mass (as in Eq. \ref{mass}); it only moves balistically in the external gravitational potential.

To summarise results of this subsection: there are two broadly defined epochs in the life of the structure.
The first epoch (coincident with the young age and the expansion phase) is characterised by the 
small dependence on the gravitational potential and (after very early time)
the large dependence on the density stratification.  The second epoch (old age or the ballistic
phase) is characterised by the large dependence on the gravitational potential and the small (or hidden)
dependence on the density stratification. In the case described in this section, the division line
lies somewhere between 30 and 50 Myr. The evolution of shells in gaseous thick discs (in
our case meaning with Gaussian scale length $\geq 300\ \mathrm{pc}$) is similar to the
evolution in the homogeneous medium, that is, the influence of stratification is negligible.
Thinner discs are different: shells experience the blow-out which leads to the energy leakage,
in-plane sizes are smaller and the shell is clearly elongated in the direction
perpendicular to the disc.

Experiments with other values would lead to similar qualitative results,
with slightly changed intervals for the young/old age of structure, or 
the value dividing thin and thick discs.

\section{Models versus observations}
\label{sec:models}

\begin{figure}
\centering
\includegraphics[angle=-90,width=0.9\linewidth]{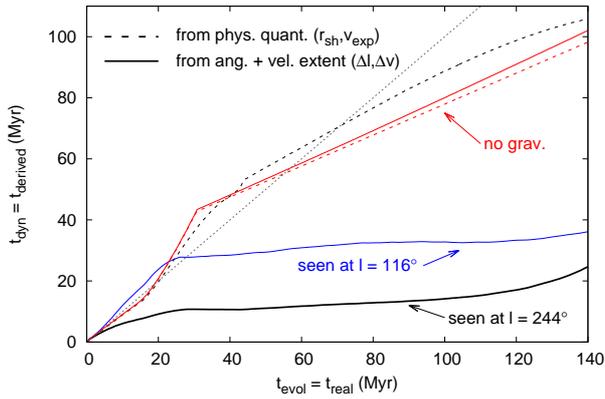}
\caption{Dynamical age $t_{\mathrm{dyn}}$ (Eq. \ref{dyntime}) as a function of the real evolutionary
  time $t_{\mathrm{evol}}$ for the basic model. Dashed lines are calculated from sizes and
  velocities taken directly from models, and solid lines are solutions derived from angular sizes and
  velocity extents corresponding to methods used in the Milky Way. Red colour is the
  model without the external gravitational potential (e.g. without the galactic
  shear). Black and blue lines correspond to the same model but viewed, at each time $t_{\mathrm{evol}}$,
  at a specified position in the Milky Way:
  at $l=244^{\circ}$ (black) and at $l=116^{\circ}$ (blue, symmetrically
  around the anticenter line). The grey dotted line is the analytical solution \citep{weaver1977}.
     }
       \label{tdyn-2met}
\end{figure}

\begin{figure*}
\centering
\includegraphics[angle=-90,width=0.45\linewidth]{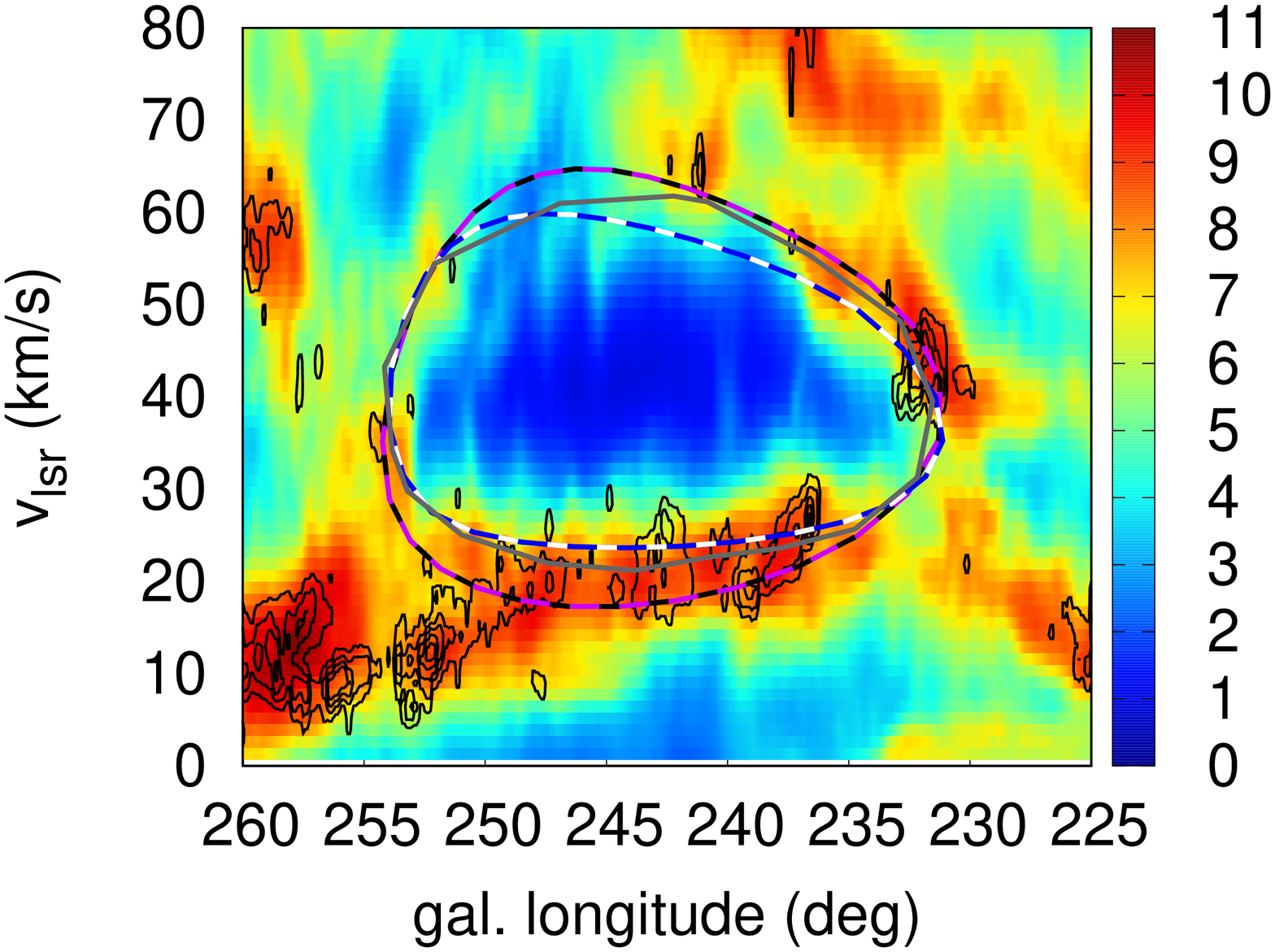}
\includegraphics[angle=-90,width=0.45\linewidth]{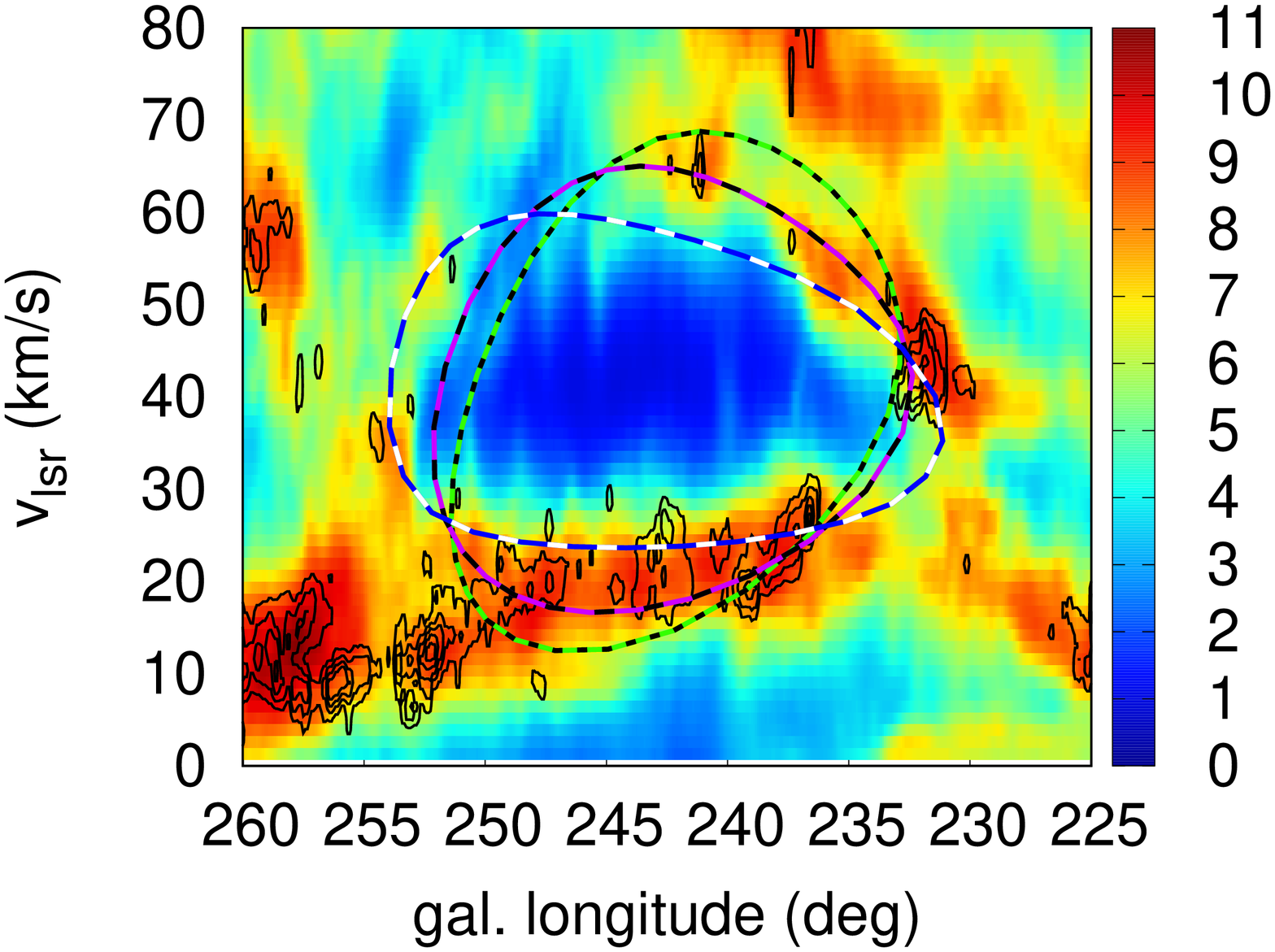}
\caption{The lv-diagram of the supershell GS242-03+37 with the outlines.
  Left: outlines of the best model
  (``basic model'' at 120 Myr, blue/white), the model with twice the energy input at 100 Myr
  (violet/black), and
  the approximate outline of the observed structure (grey line).
  Right: outlines of the best (basic) model (blue/white) and of models
  with the fixed ages (40 Myr: green/black; 80 Myr: violet/black).
  HI map is underlined; black contours show the CO emission. The colour scale is $\sqrt{T_{\mathrm{B}}}$.
  }
       \label{lv-hico-2mod}
\end{figure*}

Our aim is to compare observations of the supershell GS242-03+37 with RING models and obtain
an estimate of the energy input needed to create the shell, its age, and its position
in the Galaxy. This section first describes how we found the suitable
model and then the model itself. 

HI observations provide us with information about the brightness temperature
at two spatial coordinates and one velocity coordinate ($l$, $b$ and $v_{\mathrm{LSR}}$). We can get the same
information from numerical models and directly compare them; but looking at the beautiful image of the supershell
in Fig. \ref{gs242-lb}, with many small-scale substructures and irregularities, 
we become slightly sceptical about the reasonability, stability, and credibility
of a direct comparison of 3D datacubes with models of expanding shells in a smooth medium:
would it really be dominated by the large-scale influences
(which are the only ones we can test with our models) or not?
We believe that the main shape of the structure, resulting from the effects of the galactic
potential and the large-scale density distribution, is only partially ``smeared'' by the noise
connected to the small-scale perturbations (mostly density perturbations but also velocity ones),
and that the overall form is kept. Therefore, we
fit the outline of the simulated model to the outline of the observed structure.

As explained in the previous section, shapes of modelled shells mostly
depend, based on their age, either on the density distribution (in case of young
structures) or on the galactic potential (for old structures). From the in-plane outline
of the shell we can estimate its age and if it is young, we can then try to estimate
the density profile of the disc. 

In the case of HI observations we do not know the distance, only the radial velocity, therefore
an $xy$ shape of the supershell in the galactic plane is not explicitly known. We can however
compare the observed $lv$ (position-velocity) information to our simulations, assuming
the position of the shell and the rotation curve.
Figure \ref{lv-evolution} shows the evolution of the same models as in Fig. \ref{xy-evolution};
the centre of the model is placed at the heliocentric distance
$d_{\mathrm{hc}} = 4\ kpc$, $l = 244.5^{\circ}$ and $b = 0^{\circ}$.
We also see a very small dependence on the density stratification in these in-plane cuts, and for older ages
a large dependence of the shell $lv$-shapes on the galactic potential.

Therefore we reduce the HI observations to an $lv$ outline of the structure, which we
construct in the following way: for each $lv$ map at latitudes $b \in (-3.5^{\circ},+0.5^{\circ})$ we draw
the best outline (using pieces of walls or clumps along the perimeter and interpolating
between them) and then we calculate their average. The resulting curve is shown in Fig. \ref{lv-hico-2mod}.
As we are mostly interested in large-scale influences, the small inaccuracies are not overly important. 
The range in $b$ from which we calculated an average was chosen because neither the size of the structure
nor its shape changes significantly in this interval (as can be seen in the $bv$ map in 
Fig. \ref{gs242-lvbv}). Outside this interval the shell is diminishing and/or disappearing
from view.

\subsection{Finding the good model}
\label{sec:goodmodel}

With our models we try to fit the average measured $lv$-outline of the GS242-03+37 supershell.
We have several parameters to fit:
\begin{enumerate}
\item{Position of the structure, that is, the position of its centre. We know (from pictures)
that the centre should be located around $l=242^{\circ}$ at the heliocentric distance
$d_{\mathrm{hc}} \simeq 3.6 kpc$, but we vary these positions.
$l$ and $d_{\mathrm{hc}}$ are free parameters, not input values, and we get their
best estimates from the best fit. Since we only fit the $lv$-outline, the $b$
coordinate is not important and we do not search for it. From Fig. \ref{gs242-lvbv} it is seen
that the maximum density lies at around $b = -2^{\circ}$, because the HI disc at this longitude and
galactocentric distance is already warped from its central position $b = 0^{\circ}$}.
\item{Energy input, that is, the amount of energy delivered to the structure and the time interval
during which it is released.}
\item{The density distribution is not a free parameter. As we have shown above, the $lv$-outline
  of the shell does not depend on the stratification in the case of the thick disc, therefore
  we let the shell evolve in a homogeneous medium with 
  the density $n_0 = 0.2\ cm^{-3}$.}
\item{The age of the structure is mostly not a free parameter, unless stated otherwise. We take
  the age of the best fit for a given position, energy, and density distribution.}  
\end{enumerate}

Tested ranges for the parameters were:
$l \in (238,250)^{\circ}$;
$d_{\mathrm{hc}} \in (3,5)\ \mathrm{kpc}$;
$E_{\mathrm{0}} \in (0.1,100)\times 10^{51} \mathrm{erg}$;
$L_{\mathrm{source}} \in (0,20)\ 10^{51} \mathrm{erg\ Myr}^{-1}$;
$t_{\mathrm{energy}} \in (0,20)\ \mathrm{Myr}$;
the total energy is $E_{\mathrm{tot}} = E_0 + L_{\mathrm{source}}T_{\mathrm{energy}}$.
The density of the homogeneous medium in which the supershell evolves ($n_0$) is fixed,
because it is an estimate of the observed average density at the position of the shell.
As we have shown in the previous section, the evolution of the shell, even for more complex
models, depends on the ratio $L/n_0$, so in cases where our chosen density is shown to be too
low or high (0.3 instead of 0.2 e.g.), we can easily rescale the fit.

The model $lv$-outline from the RING code (as in Figs. \ref{lv-evolution} or \ref{lv-hico-2mod})
is compared to the measured $lv$-outline. From the previously given net of parameters we find the best,
that is, most acceptable, fit with values:
$l = 244.5^{\circ}$, $d_{\mathrm{hc}} = 4.0\ \mathrm{kpc}$,
$L_{\mathrm{source}} = 0.8\times 10^{51} \mathrm{ergMyr^{-1}}$, $t_{\mathrm{energy}} = 15\ \mathrm{Myr}$.
The total energy of this model is $E_{\mathrm{tot}} = 13 \times 10^{51} erg$.
The corresponding age of this fit is $\tau = 120\ Myr$. This is what we refer
to in the following text and in the description of figures as the ``basic model''.

The fitted heliocentric distance of the shell is larger than the distance from a simple estimate based
on the  `central' velocity of the structure, as given in section \ref{sec:gs242}. This
simple estimate corresponding to the radial velocity of $38\ kms^{-1}$ is $d_{\mathrm{hc}} = 3.6\ kpc$,
but best fits to observations favour the larger distance of $d_{\mathrm{hc}} = 4\ kpc$. Given
the fact that the shell is large and its shape and velocity field are distorted by the differential rotation,
it is not surprising that there is a small difference between these two estimates. 

Figure \ref{lv-evolution} shows the $lv$ evolution of the basic model. Figure \ref{lv-hico-2mod}
shows the best fit of the basic model at $\tau = 120\ Myr$ and also the best fit of the model, which has
the twice larger energy input ($L_{\mathrm{source}} = 1.6\times 10^{51} \mathrm{ergMyr^{-1}}$);
the best age in this case is 100 Myr.
This comparison also illustrates how the best fit should be treated: as a representative of
  a family of suitable fits. Slightly changing one parameter of the fit would produce reasonable results;
  changing it too much damages the fit. This `uncertainty' is mostly connected to uncertainties and free will
  in the definition of observed shape and size of the shell, and to a lesser degree to inherent degenerations
of parameters, like the $E_0/n_0$ degeneracy discussed previously.

Our model predicts that the shell was supersonically expanding for the first $\sim$ 30 Myr
of its evolution and now it propagates as a sound
wave with the velocity of $\sim 7\ kms^{-1}$. Its large observed velocity extent
is mostly caused by the galactic rotation. This agrees with the result derived by \citet{mcclure2006}.

An important aspect of our best fit is the old age of the structure, seemingly inconsistent
with the simple age estimate of Eq. \ref{dyntime} ($\sim 20\ \mathrm{Myr}$).
This young age is obviously very different from
the age derived from the fit. The main reason for the discrepancy is the influence
  of the galactic gravitational field, which causes
the galactic shear. Figure \ref{tdyn-2met} shows how the
dynamical age of the simulated structure derived using Eq. \ref{dyntime} changes with time.
For $r_{\mathrm{sh}}$ and $v_{\mathrm{exp}}$ we use
  \begin{enumerate}
  \item{$r_{\mathrm{sh}}$ and $v_{\mathrm{exp}}$ ,which are calculated as the average values of the model in the galactic
    plane (dashed line in the Fig. \ref{tdyn-2met}), this method is available only for models, not for real
    structures observed in the Milky Way;}
    \item{$r_{\mathrm{sh}}$ and $v_{\mathrm{exp}}$ , which are derived from measured dimensions $\Delta l$ and $\Delta v$ of the
      shells and from the assumed distance to the shell (solid line in the Fig. \ref{tdyn-2met}),
      this is the way used for observed structures.}
  \end{enumerate}  
Calculations are done for the shell located at $l_0=244.5^{\circ}$, that is, for the real position
of the supershell (black lines in the Fig. \ref{tdyn-2met}) and also for the symmetrically positioned
shell at $l_0=115.5^{\circ}$ at the
same galactocentric distance (the blue line).  Method~1, which uses real radii and velocities of the
model, gives the same values for both positions (i.e. for all positions, which differ only in the
galactic longitude). To see the effect of the diffential
rotation we overplot the model calculated without the external gravitational field in Fig.11 (red lines). The analytical
solution of \citet{weaver1977} is shown as the green line. The energy input to the shell is stopped
at 15 Myr, which leads to deviations from the analytical solution even for the simple shells evolving
without the external gravitational field. These deviations are not substantial, however, and the simple estimate
gives a reasonable value for the dynamical age. Even method~1, using the average values of
radius and expansion velocity of the shell, works reasonable well (unfortunately, for real observed structures
in the Milky Way
we cannot use it). Method~2, using the projected angular and velocity dimensions, gives
incorrect values, which are dependent on the direction in which we observe. In our case simulations
predict that the dynamical age derived from the $\Delta l$ and $\Delta v$ is nearly constant with time
(and therefore not usable at all). The main reason for this behaviour is the falsely higher expansion velocity,
which is due to the galactic shear, not due to an expansion. The derived radius is also dependent on the
direction in which we observe (the difference between shells at $l_0=244.5^{\circ}$ and $l_0=115.5^{\circ}$),
because it decides if we observe along the major or minor axis of the shell (see Fig. \ref{xy-evolution}
for differences in sizes).
Our modelled shell, that is, the basic model at the position $l_0=244.5^{\circ}$, is observed
not directly along the minor axis, but nearly so (as is shown in Fig. \ref{wall-clusters-names}
below), therefore its angular dimension almost corresponds to its largest size.

\subsection{Models with the fixed age of the shell}
\label{sec:fixedage}

The models described above have several free parameters: $l$ and $d_{\mathrm{hc}}$ describing the
position of the centre of the supershell, $L_{\mathrm{source}}$ and $t_{\mathrm{energy}}$ for energy requirements,
and the age of the structure $\tau$, which is a result of the search for the best fit.
But $\tau$ can also be fixed. To see results for younger structures we made calculations with the
same free parameters as above ($l$, $d_{\mathrm{hc}}$, $L_{\mathrm{source}}$, $t_{\mathrm{energy}}$), but this time fixing the evolutionary time
(i.e. the age of the structure). In Fig. \ref{lv-hico-2mod} (right panel) we show the best results for shells
with ages 40 and 80 Myr. The total energies of these models are $1.2\times10^{53}\mathrm{erg}$ for
$\tau = 40\ \mathrm{Myr}$ and $3.0\times10^{52}\mathrm{erg}$ for $\tau = 80\ \mathrm{Myr}$.

The 40 Myr model does not give a good fit. To reach such a large size in a relatively short time the structure needs
a lot of energy and therefore its expansion velocity is large.  The combination of
demands on the size and on the velocity does not give a satisfactory result: even the best
fit shown in Fig. \ref{lv-hico-2mod} underestimates the angular size and overestimates the velocity
extent.

A fixed age of 80 Myr is a better condition for the shell. However, the agreement between the model and the
observations is still not as good as for our basic model at 120 Myr, especially around $l = 252^{\circ}$. The 80 Myr
model perhaps\ also predicts the $v$-extent to be too large --- reaching lower radial velocities than
observed --- but it fits observations better than the younger model and might be acceptable.

\subsection{Is our best model physically acceptable?}

  As stated above, the most surprising result of our modelling and fitting is the large age of the
  supershell. The simple dynamical age of the shell based on Eq. \ref{dyntime}
  is 20 Myr (assuming that the velocity interval in which we observe the shell corresponds to the
  real expansion velocity; if the shell does not expand supersonically, but is already
  stalled, its dynamic age is therefore $\sim$ 70 Myr). 
  Our fit is 80 - 120 Myr, with the best fit corresponding to 120 Myr. The major issue
  is not only the simple discrepancy between these numbers,
  but the large age itself. Can a supershell survive for such a long time?

  There are several ways how the shell can disappear: by dissolving into the ISM, by being
  destroyed in a collision with a different structure, by twisting too much by the differential
  rotation, and by the passage of the spiral arm. We discuss these separately below.

\subsubsection{Dissolution into the ISM}

The shell can simply dissolve after it ceases to expand supersonically
\citep[referred
to in][as the disintegration of their stalled shells]{oey1997}.
This dissolution can take place either via dissollution of the dense walls of the shell or via the flow of ambient
medium  from the unperturbed medium outside the shell to its low-density interior.
Efficiency of the wall dissolution and the hole refilling (and probably also the steepness
in density contrast between a shell interior and its walls) depend on the compactness or
fragmentation of dense walls.

Shells are subject to different fragmentation processes. \citet{dove2000} discuss the Rayleigh-Taylor
and gravitational instabilities and they conclude that fragmentation of a shell is likely to start
about 30 Myr after the beginning of energy injection, at the time when the supershell
collects enough mass, when it cools, and when it increases the density contrast between the shell wall and
the hot gas in the interior. For our model, this fragmentation time is also the time when
the expansion of the shell starts to be subsonic. After that point, fragments expand ballistically
in the gravitational field of the Galaxy and are not influenced by each other or by the low-density
hole swept-up in the ISM. Instabilities and fragmentation of pressure confined in self-gravitating layers
is also discussed by \citet{dinnbier2017} using results of hydrodynamical simulations.

Clumps in walls of GS242-03+37 are visible in Figs. \ref{gs242-lb}, \ref{gs242-lvbv}, and \ref{tbprofile}, 
but the structure --- both dense walls and the low-density hole --- keeps its coherence. The lifetime of
individual fragments in walls depends on their size, temperature, and density.
Some less dense and hotter fragments probably dissolve and contribute to the density of the
interior gas. Remaining fragments are observed in HI and CO at the rim of the GS242-03+37.
The volume density inside the fragments in the wall may be substantially larger than
the density of the ambient medium $n_0$.
  The self-gravity of these denser fragments probably helps in maintaining the high contrast between
  walls and an interior by slowing down the process of evaporation from the clumps and their
  subsequent dissolution.

    A rough estimate of the importance of the self gravity is the ratio between the mass of the fragment
    and its virial mass: if it exceeds 1 (or, given all the uncertainties, close to one), the self gravity
    is important. This ratio is
 \begin{equation}
   {M_{\mathrm{fr}} \over M_{\mathrm{vir}}} \sim
   {1 \over 50} {\Delta s_{\mathrm{fr}} N_{\mathrm{fr,20}} \over \sigma^2}
  \label{rat_fragvir}
 ,\end{equation}
 where $\Delta s_{\mathrm{fr}}$ is the size of the fragment in pc, $N_{\mathrm{fr,20}}$ its column density
 in $10^{20} \mathrm{cm}^{-2}$ and $\sigma$ its velocity dispersion in $\mathrm{kms}^{-1}$. The average
 column density of the supershell is $(1-2)\times 10^{20} \mathrm{cm}^{-2}$, calculated as the total
 mass divided by the surface (from which $N_{\mathrm{fr}} = r_{\mathrm{sh}} n_0 / 3$). For fragments with
 sizes of about 100 pc, comparable with the thickness of the shell, and for the purely atomic gas
 with $\sigma \sim 10\ \mathrm{kms^{-1}}$ the ratio is very small ($\sim 0.02$). However, if the
 fragment is cold, its velocity dispersion can decrease to $2 - 3\ \mathrm{kms}^{-1}$.
 Consequently the ratio ${M_{\mathrm{fr}} / M_{\mathrm{vir}}}$ is much higher and therefore the self-gravity
 is more important.

  Refilling of the shell interior from outside can start when the shell stops the supersonic
  expansion and slows down to the speed of sound in the ambient medium. A hole with a diameter
  of 1 kpc can be refilled, perhaps not completely but significantly, in about 50 Myr (assuming
  the speed of sound $c_{\mathrm{s}} = 10\ \mathrm{kms}^{-1}$). However, GS242-03+37 is filled but not
  significantly (the ratio between the derived interior and outside volume density is about 1/4;
  see Fig. \ref{HIdensmap}). Are there mechanisms that might prolong the refilling? 
  First, if the fragments are cold, either molecular or cold atomic,
  the ambient gas can only flow to the interior through gaps in fragmented walls
  since the cold fragments interact with the warmer and supersonically moving gas
  flowing from outside. If the majority of the wall is made
  from cold dense fragments, the refilling is substantially lowered.
  This process depends heavily on fragments' filling factor and their compactness;
    without the more detailed analysis of fragments in both HI and CO it is not possible
    to give an accurate estimate of its influence. Such an analysis is beyond the scope
  of the current paper and would perhaps need the CO data with a higher spatial resolution.

  Second, the shell is extended
  in the $z$-direction. We estimate its current $b$ size as $12^{\circ}$, meaning more than 800 pc,
  with the possibility of being at least partially open \citep[as given by][]{mcclure2006}
  to the halo. This means that a part of the ambient gas, which mostly originates from the densest
  parts of the galactic disc, flows to higher |z| locations, where the intrinsic refilling is negligible
  due to the low ambient density, or eventually can even escape from the supershell. This again
  slows down the rate at which the hole is refilled.

  Taken together, these findings suggest that we need to assume that the evaporation from walls and the refilling by
  the ambient medium was slowed down by some or all of the processes described above, or due to some
  other mechanism, and therefore we still see the shell with a high contrast after about 80 Myr of subsonic
  expansion. 
  An alternative, which we do not advocate, but which cannot be ruled out
  by our models, is a slightly lower age of the structure.  As explained above (in section \ref{sec:fixedage},
  see also Fig. \ref{lv-hico-2mod}), there are some not particularly good but acceptable
  fits which predict lower ages (by 80 Myr or more) than the basic model, in which the subsonic expansion
  would last for only 40 Myr.
  To resolve the problem of the long-term existence of GS242-03+37 a more detailed identification
    of fragments in HI and CO 
  and a full hydrodynamical model would be needed.

\subsubsection{Collision of shells}
  
The shell can be destroyed in a collision with another shell. This is not a probable
scenario for GS242-03+37 since it is so big and it lies in the outer Milky Way where the density
of shells and similar structures is low, contrary to the inner Milky Way. Also, merging
of shells does not need to destroy them, it can only create a larger shell
\citep[see][where merging of small shells is shown to be an important mechanism for creating supershells]{krause2015}.

\subsubsection{Differential rotation}

The differential rotation can twist the shell so much that it ceases to be distinguishable. This is what
we model and test with our simulations.
The eccentricity of the shell, that is, the ratio between the minor and major semi-axes, grows
  with time. The shell at the galactocentric radius of the Sun (8.5 kpc) and with our adopted
  rotation curve, reaches the eccentricity of 0.5 at 46 Myr and the eccentricity
  of 0.1 (i.e. very distorted) at 135 Myr. The shell at the galactocentric distance of 5 kpc
  evolves faster, reaching eccentricities 0.1 and 0.5 at 28 and 83 Myr, respectively. On the other hand,
  the shell at 11 kpc distance evolves slower; the adequate times are 63 and 180 Myr. The eccentricity
  of the shell at the predicted age (120 Myr) is 0.3, very elongated but not yet destroyed.

  The intrinsic eccentricity of GS242-03+37, which would be most visible along the $z$-direction
  from outside the Milky Way, is difficult to verify with observations, since our position relative to
  this structure does not enable us to see its real shape in the galactic plane.

\subsubsection{Spiral arms}
  
Shells can be destroyed by the passage of the spiral arm. For a discussion of the effects
of the spiral structure on shells see for example \citet{mcclure2002}. Generally, this
effect depends on the frequency with which the shell meets the spiral arm and this frequency depends
on the position in the galaxy and on the spiral pattern speed. Close to the corotation radius this effect
is not important. The pattern speed and the location of the corotation in the Galaxy are not
precisely known \citep[for a review of values of pattern speeds $\omega_p$ in the Milky Way, see][which gives
limits $\omega_p \in (17,28)\ \mathrm{kms}^{-1}\mathrm{kpc}^{-1}$]{gerhard2011}.
Our supershell is located at $R \simeq 11\ \mathrm{kpc}$ ($R/R_{\circ} \simeq 1.3$) and if it should
lie close to the corotation, the required pattern speed should be around $20\ \mathrm{kms}^{-1}\mathrm{kpc}^{-1}$,
compatible with cited limits.

We can estimate limits on the pattern speed by requiring that
  the volume of the supershell not be disturbed by the passage of the spiral arm
  for a sufficiently long time. Using our adopted rotation curve \citep{wouterloot1990}, the four-arm spiral pattern,
  the centre of the shell (at the galactocentric distance 10.8 kpc $\pm$ 1 kpc), and
  the required interval between subsequent spiral arm passages of $\geq 200$ Myr, we obtain the condition that
  $\omega_p \in (15,26)\ \mathrm{kms}^{-1}\mathrm{kpc}^{-1}$ (the result for the flat rotation
  curve is essentially the same). The `most suitable' pattern speed (with the largest
  interval between passages) is $\omega_p = 19\ \mathrm{kms}^{-1}\mathrm{kpc}^{-1}$.

\subsection{Old shells in external galaxies}

  From studies of external galaxies we know that large and old HI structures exist.
  \citet{bagetakos+2011} identify HI shells in 20 galaxies (11 spirals and 8 dwarfs)
  and 7 of them have shells with ages $\geq$ 100 Myr (4 spirals and 3 dwarfs). Several
  of these galaxies contain more than one such structure.  
  The shell ages are derived from the ratio of $r_{\mathrm{sh}}$ and $v_{\mathrm{exp}}$, i.e. as in
  Eq. \ref{dyntime}, except they use the coefficient 0.978 instead of 0.5 and therefore their ages
  are higher (the age of the supershell GS242-03+37 using the Bagetakos formula would be 50 Myr
  under the assumption that what we observe is the real expanding velocity, or 140 Myr for the
  subsonic expansion).
  
  None of these old supershells in external galaxies
  are supersonically expanding, all are large (usually around 1500 pc in diameter but
  some even larger than 2000 kpc).  The majority of these old shells
  are found inside the $R_{25}$ isophotal radius, some even at radii $< 0.5\ R_{25}$.
  Their eccentricities (ratios between minor and major axes) vary between 0.4 and 0.9,
  with 0.7 being the average value. 
  
  The large dimensions and age of the supershell GS242-03+37 are therefore fully consistent
  with the above structures. The difference is the observing point: in the case of GS242-03+47
  we observe the shell `edge-on', while the mentioned shells in external galaxies are
  seen more `face-on'. This results in a different contribution of motions parallel and
  perpendicular to the galactic plane to the observed expansion velocity of structures.
  The `edge-on' view enables observations of the in-plane motions, while the fully `face-on'
  view is sensitive only to motions perpendicular to the disc.

\subsection{Energy requirements}

An interesting consequence of the old age of the supershell is the relatively low energy
needed to create it. This is caused by the fact that the size of the structure and its expansion velocity
are partially caused by the galactic shear and not by the energy input to the shell.
Our best fit (13 $E_{\mathrm{SN}}$) is probably a lower limit, 1) because we do not
take into account any radiative losses or the leakage of energy from the shell interior
during the supersonic expansion phase and 2) because
the evolution takes place in the homogeneous medium, which, for a given energy input,
gives maximum sizes of the shell in the plane of the Galaxy. The evolution in a stratified
disc is connected with a prolongation in the direction perpendicular to the disc
and an eventual loss of energy to the halo (but since the supershell is located in
a rather thick disc, this loss is not extreme, i.e. not the complete blow-out,
as we see e.g. in Figs. \ref{xy-evolution} and \ref{xz-evolution}, and discuss in the previous section).
It is possible that a part of the energy released during the formation of the
  supershell leaked from the hot bubble and created the high latitude caps mentioned in
  \citet{mcclure2006}. This energy loss is not taken account of in our analysis.
Therefore it is natural that our energy estimates are much lower than previous ones
  \citep{heiles1979, mcclure2006}. 

  In our model we disregard the cooling from the shell interior. If the cooling were high,
  it would increase the energy requirement and perhaps decrease fragmentation times. However,
  due to the low density inside the shell ($n_{\mathrm{HI}} < 0.1 \mathrm{cm}^{-3}$), cooling
  is inefficient. During the initial 30 Myr of its supersonic expansion, its influence is
  very small. Even if this influence were to become more important later, when the HI hole is gradually refilled
  from higher-density walls, the motion of the fragmented shell in the gravitational field of Galaxy would
  no longer be influenced by this medium.   

  Given all the uncertainties and approximations, we still refer to GS242-03+37 as a supershell, even though
  our basic model suggests its energy is lower than
the minimum value of the energy of the supershell \citep[$3\times 10^{52}\ \mathrm{erg}$, ][]{heiles1979}.

\section{Distribution of clusters relative to GS242-03+37}
\label{sec:clusters}

\begin{figure}
\centering
\includegraphics[angle=-90,width=0.9\linewidth]{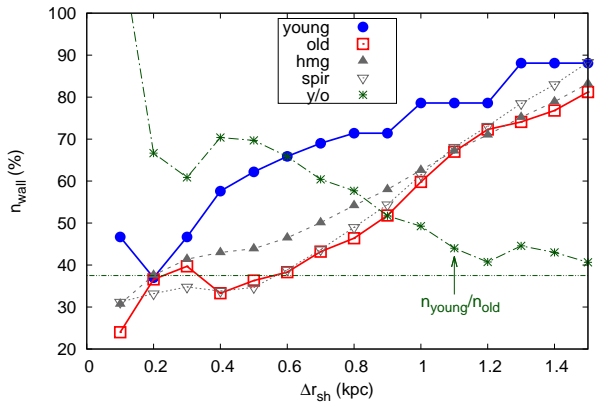}
    \caption{Number of clusters as a function of the thickness of the shell.
      Young (< 120 Myr) clusters are shown in blue, old clusters in red.
      Grey lines are randomly generated samples, with the homogeneous
      (filled) and spiral distance-dependent (empty) probability distribution.
      The green line shows the ratio between the number of young and old clusters
      in the wall, the value 37.5 \% is the ratio between the total numbers
      of young and old clusters.
    }
       \label{nwall-deltad}
\end{figure}

We compare the distribution of clusters to the distribution of gas and the supershell
GS242-03+37. We use the RING model (the best fit as described above), and not 
the HI data directly, because HI observations
give us $lbv$ coordinates (i.e. galactic longitude and latitude and a radial velocity) while for 
clusters we usually have $lbd$ coordinates (i.e. galactic longitude and latitude and a heliocentric
distance). In models, we have all coordinates; $lbv$ were used for the comparison with HI observations
while $lbd$ (or $xyz$, which is equivalent) are used for comparison with clusters.
Positions of clusters were not a parameter in the model fitting described in the previous
  section. The model of the supershell serves as the instrument recalculating radial
velocities into heliocentric distances and vice versa.

We try to test the correlation between the HI supershell GS242-03+37 and positions of star
clusters by calculating the number of clusters lying in the wall of the supershell.
The correlation between the HI supershell and star clusters, if any,
should exist for clusters younger than the supershell and should not exist for older
clusters. Therefore we separate clusters into two categories: young clusters
($\tau < 120\ Myr$; which might be influenced, or indeed triggered,  by the presence
of the supershell) and old clusters ($\tau > 120\ Myr$; which should not be influenced
by the supershell at all, since they are older than the structure).

At first glance, it is not completely obvious why the cluster with the age of $\sim$ 100 Myr
should still lie in the wall of the shell, even if it originated there. It is reasonable to assume
that a newly born cluster inherits the velocity of the gaseous material from which 
it was born. The expansion velocity of the HI shell is slowing down with time. Therefore
the trajectory of the cluster is (theoretically) different from the trajectory of the wall. 
But this applies mainly to clusters born in an early phase of evolution, where
the expansion velocity of the shell is high, that is, when the shell was younger than 20 Myr or so.
If clusters are born after this period, i.e. when the shell is older than about 20 Myr,
the difference in velocities (and thus in positions) of the cluster and the wall is small
and the cluster still resides in the vicinity of the wall.

\subsection{Number of clusters in the wall of the supershell} 

From our numerical simulations we know the position of the dense shell (in $xyz$ or $lbd$ space)
but we do not know its thickness as our simulations assume that it is infinitesimally thin.
Let us define the thickness of the shell wall as $2 \times \Delta r_{\mathrm{sh}}$.
A cluster resides in the wall if its closest distance to the shell,
$\Delta d_{\mathrm{c-w}}$, is less than $\Delta r_{\mathrm{sh}}$: $\Delta d_{\mathrm{c-w}} \le \Delta r_{\mathrm{sh}}$; this way we explore the concentration of clusters to the wall.

Figure \ref{nwall-deltad} shows numbers for young (age $<$ 120 Myr) and old
(age $>$ 120 Myr) clusters as a function of $\Delta r_{\mathrm{sh}}$.
The studied area is $l \in (230,255)^{\circ}$, $b \in (-3,+3)^{\circ}$
and  $d_{hc} \in (2.5,5.4)$ kpc (in the subsection we have trimmed the area so as not to
deal with a lot of space far from the supershell).
Since the absolute number of young and old clusters in the area differs (42 or 112, respectively), we use
the normalised value, that is, the absolute value divided by the total number of young or old clusters.

Consistently larger number of young clusters are shown to be sitting in the wall, except for
very thin walls, where any error in the distance determination, either of the cluster or of the
wall, plays a large role; or very thick walls, where the
substantial part of the area is covered by walls and hence the distinction wall/not-wall
loses sense.

To compare the calculated dependence of the number of observed clusters in the wall on the thickness
of the wall, we have also calculated the expected number of clusters for two
distributions: the `homogeneous' distribution of clusters in the area, where
clusters are homogeneously distributed in both $l$ and $d$ directions
($b$ direction does not influence the analysis); and the `spiral' distribution,
which copies the distribution of all clusters with the distance shown in Fig. \ref{clusters-dhc}.

In both cases, these randomly generated samples have a smaller
number of clusters in the wall than corresponds to the observed distribution of young
clusters, but very similar to the distribution of old clusters. The spiral distribution is more
similar to the distribution of old clusters than the homogeneous one, which is understandable, since it
reflects the distribution of clusters with the distance. Figure \ref{nwall-deltad} also
shows the ratio between the number of young (< 120 Myr) and old (> 120 Myr)
clusters in the wall (in percents); the value 37.5 corresponds to the ratio between the total
number of young clusters in the area (42) and the total number of old clusters (112).

The conclusion is simple: the fraction of young clusters lying in the wall of
the supershell is larger than for old clusters or for randomly generated cluster samples.

\subsection{Individual clusters associated with the shell}

\begin{figure}
  \centering
  \includegraphics[angle=-90,width=0.9\linewidth]{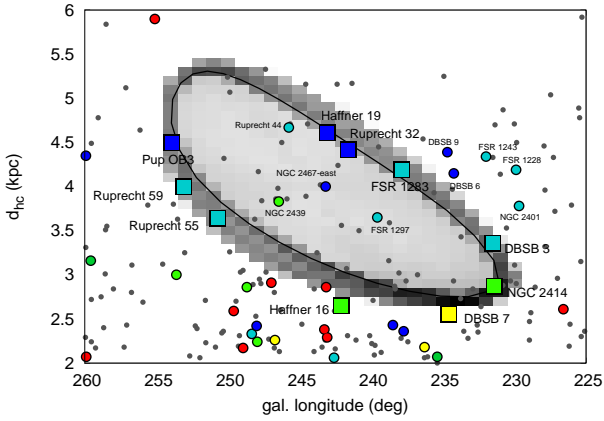}
  \caption{The $ld$ diagram shows the model of the supershell projected to the
    longitude-heliocentric distance plane (the black outline and the grey map, the intensity corresponds
    to the gas mass) together with positions of star clusters. Ages of clusters are colour-coded:
    < 5 Myr (blue), 5-10 Myr (cyan), 10-15 Myr (darker green, only 2 clusters in the image),
    15-20 Myr (light green), 20-40 Myr (yellow), 40-120 Myr (red), and > 120 Myr (dark grey). 
    Clusters most probably originating in the shell are denoted as large squares and their
    names are given, as well as names of some other young clusters.
  }
  \label{wall-clusters-names}
\end{figure}

\begin{figure}
\centering
\includegraphics[angle=-90,width=0.9\linewidth]{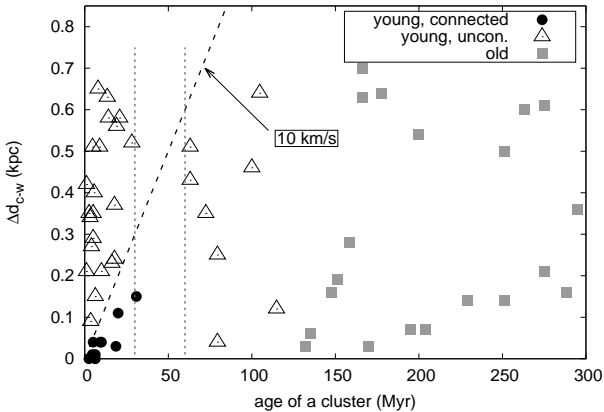}
\caption{Distance of the star clusters from the wall ($\Delta d_{\mathrm{c-w}}$)
  as a function of age. Clusters, most probably created in the wall of the supershell
  are shown as black circles, other young clusters (age $<$ 120 Myr) are empty triangles.
  Old clusters (age $>$ 120 Myr) are grey squares. The dashed line shows the distance
  travelled at a speed of 10 $\mathrm{kms}^{-1}$ during the age of the cluster. Dotted
  line delineate ages of 30 and 60 Myr.
    }
       \label{dring-age}
\end{figure}

We have shown in the previous subsection that young (age $ < 120$ Myr) clusters tend to be associated
with the wall of the supershell GS242-03+37. This association is statistical,
but we would like to know which clusters came into existence in the wall, that is,
were triggered by the supershell.
It is difficult to tell for sure which individual clusters originated in the wall,
but we can estimate the probability of the association and choose the most probable clusters.

We have three criteria by which we evaluate if an individual cluster
could be connected to the HI supershell:
\begin{enumerate}
\item{The position in the wall, that is, the distance from the
    position of the shell, $d_{c-w}$,  is very small (less than e.g. 100 pc).
    This condition favours very young clusters, since they do not have sufficient
    time to move away from the place of their birth. Older clusters, even if they
    originate in the wall, may be further away simply because they naturally move
    away.}
\item{It is easy to move to the current position, that is, $d_{c-w}$ is such
  that it could be travelled during the lifetime of the cluster
  with a reasonable velocity (e.g. less than $10-20\ kms^{-1}$). It favours older clusters since
  they have lots of time to move far from the wall.}
\item{The radial velocity of the cluster is
  similar to the radial velocity of the nearest part of the shell. The difference should not be too large,
  it is basically the difference in expansion velocities at the time of the cluster creation and now.}
\end{enumerate}

We have sorted out nine clusters, which satisfy these criteria best:
Haffner 19, Haffner 16, Ruprecht 32, NGC 2414, Ruprecht 59, FSR 1283, DBSB 3, Ruprecht 55,
and DBSB 7. All these clusters lie very close to the shell.
Seven out of them are at distances smaller than 50 pc from the shell; the remaining two
(Haffner 16 and DBSB 7) are slightly further away (100 - 150 pc), but since their ages
are 20 - 30 Myr, they could easily have travelled this distance during their lifetimes.
Only three other clusters are found to be less than 150 pc from the shell.
One of them is a young cluster Ruprecht 44: it was not chosen as a suitable candidate
because 1) it is young and therefore would need
relatively high velocity to reach its current position, and 2) it lies inside the shell.
Further discussion about this cluster can be found in the subsection ``Strange case of NGC 2467-east''.
Another young cluster, DBSB 6, lies outside the wall, but it would need a high extra velocity
to reach its current position ($> 25\ \mathrm{kms}^{-1}$), so it was disregarded for this reason.
One older cluster, Haffner 18, lies very close to the wall (and due to its age has no problem in reaching
its current destination), but there is a high discrepancy between its distance (and predicted radial
velocity) and the observed radial velocity; for this reason it was not chosen.
All other clusters in the area containing the supershell are further from the shell.

At Fig. \ref{dring-age} we plot the star cluster ages versus their distances to the wall
($\Delta d_{\mathrm{c-w}}$). Ten clusters (the nine listed above and Pup OB3 --- see below), which are
most probably connected to the shell, are
shown as black circles. The older ones of these reside close to the tip of the supershell twisted
by the differential rotation, others are in walls/sides. There are no clusters with ages between
30 and 60 Myr in the supershell region. This time interval, due to $z$-oscillations
perpendicular to the disc, corresponds to the period of a lower density in shell walls
(see Fig. \ref{pot-evolution}, right panel).

Table \ref{table:1} contains all relevant information (coordinates, ages, distances)
about the star clusters younger than 120 Myr in the supershell  GS242-03+37 region
(displayed in Fig. \ref{dring-age}). The table also shows distances of clusters to the
wall of the supershell ($d_{\mathrm{c-w}}$) and highlights the clusters associated with the
HI structure.
There are only three clusters with ages > 80 Myr, as also seen in Fig. \ref{dring-age}.
  Therefore, if the supershell is younger than 120 Myr, for example 80 Myr, which is our lower
  limit on the age of the supershell, results derived in this section would still be valid, because the analysis would be only
  slightly changed.

\subsection{Young cluster in the wall: a case of Pup OB3}
\begin{figure*}
\centering
\includegraphics[angle=-90,width=0.45\linewidth]{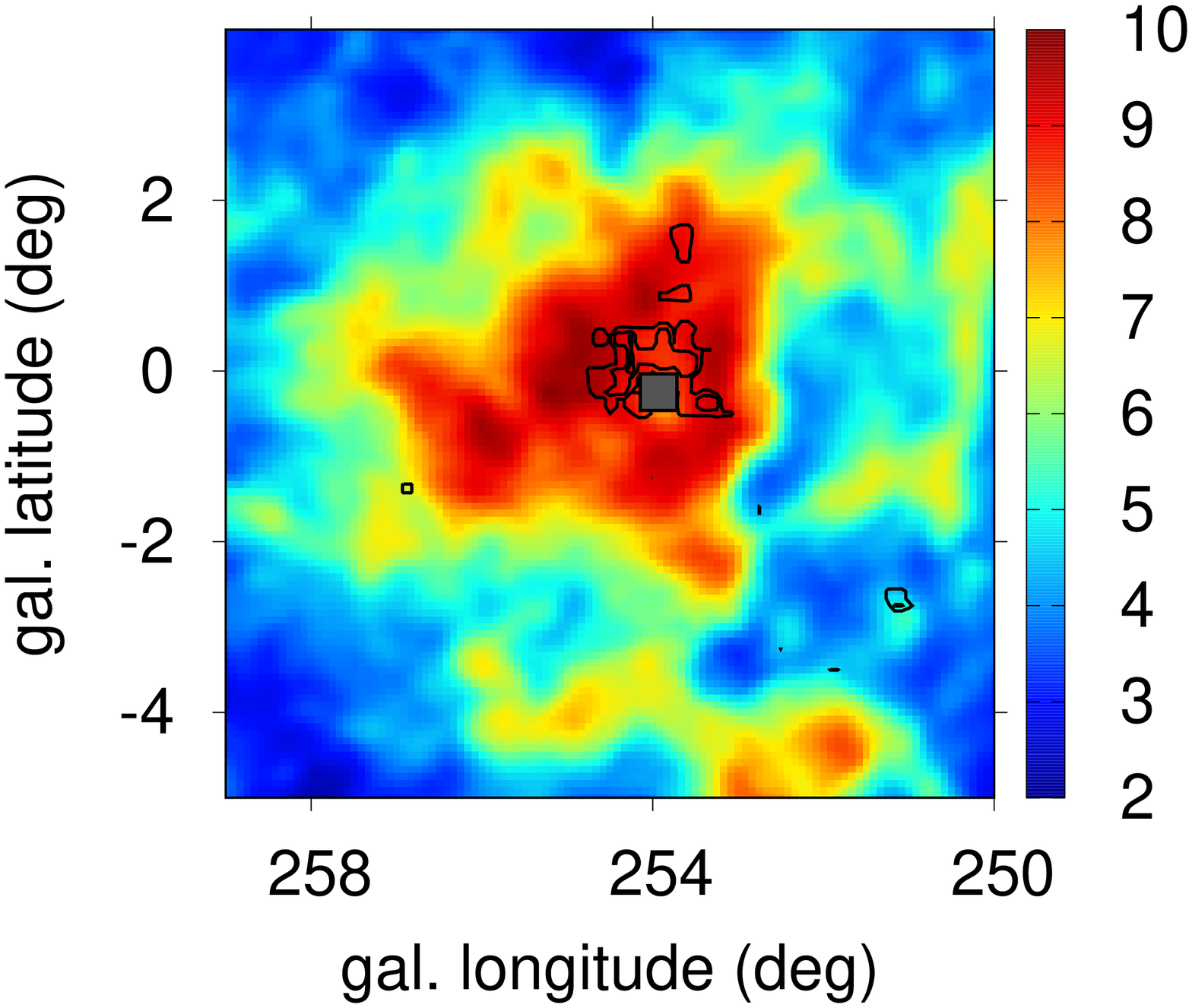}
\includegraphics[angle=-90,width=0.45\linewidth]{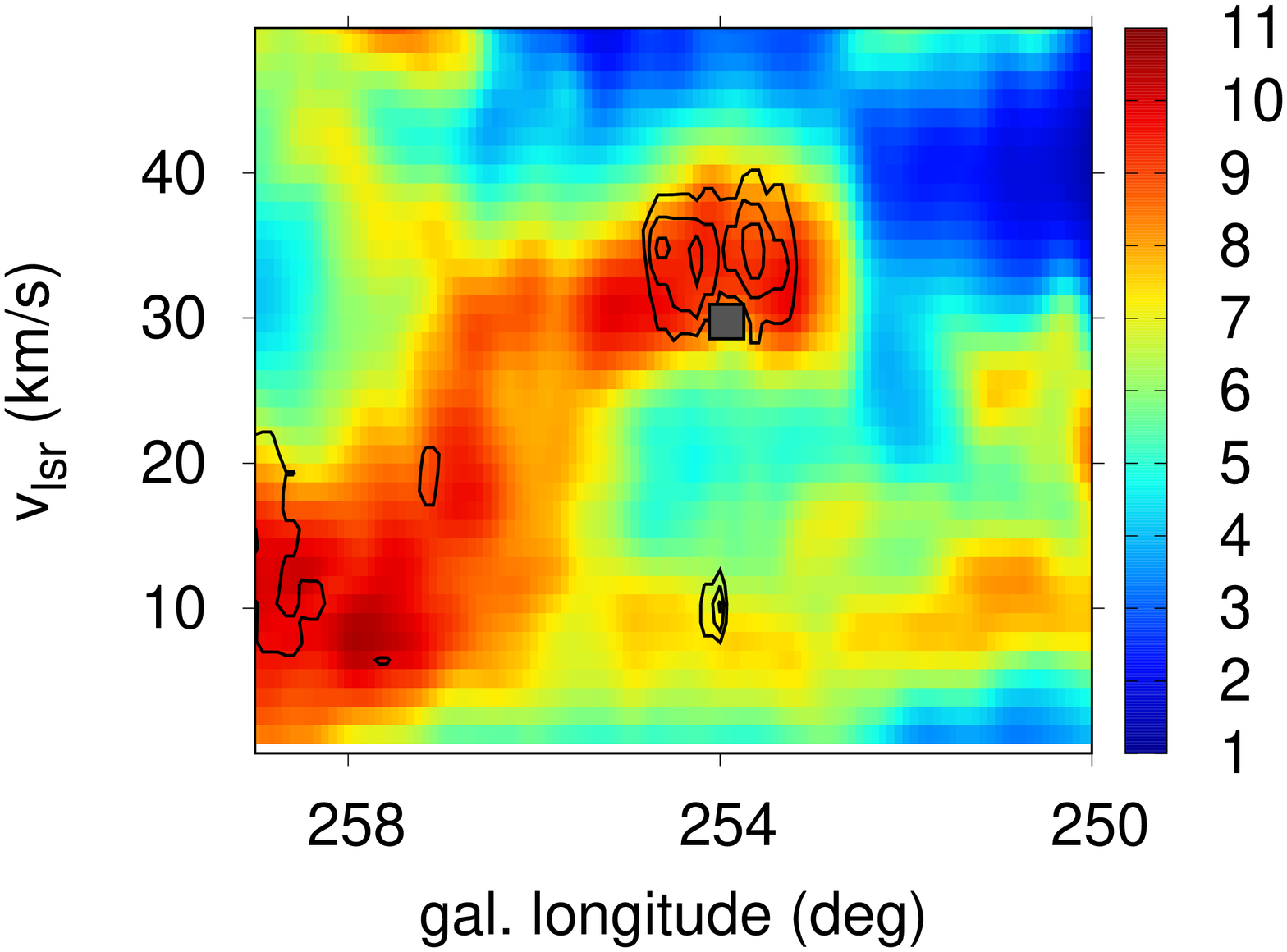}
\caption{Pup OB3 and the surrounding ISM. The map corresponds to the HI emission,
  contours to the CO emission. The grey square denotes the position of Pup OB3.
  The left panel is the channel map (velocity interval: $<28.0,32.0>\ kms^{-1}$),
  the right panel is the $lv$-diagram (in the $b$ interval of $<-0.5^{\circ},0.0^{\circ}>$).
     }
       \label{pupob3-lblv}
\end{figure*}

In all analyses presented here, cluster data (positions, distances, ages
and radial velocities) were taken from the catalogue of \cite{kharchenko+2013}, even though
in some cases other authors give alternative values of distances and ages. As  explained
above however, we prefer the homogeneity of the catalogue to the randomness of many sources.

There is, nevertheless, one exception, the cluster Pup OB3. This is a young OB association (2-3 Myr)
at the $v_{lsr} = 30\ kms^{-1}$. This cluster still seems to be clearly associated with the gas in its
surroundings (see Fig. \ref{pupob3-lblv}); it might be associated with the HII region
Gum 10 = RCW 19 \citep{avedisova2002}. The radial velocity of RCW 19 is
$v_{lsr} = 29.5\ \mathrm{kms}^{-1}$, with the dispersion $\sigma = 12.9\ \mathrm{kms}^{-1}$  \citep{g+g1970}.
Based on the assumption of the physical connection between
the gas observed at $v_{lsr} = 30\ kms^{-1}$ (which lies in the wall of the supershell) and the Pup OB3,
the distance to the association should be around 4.5 kpc, not the 1.7 kpc as given in the
catalogue. Since the association between the cluster and the gas is tight - as expected for
such a young object - we propose that Pup OB3 is lying in the wall of the supershell and add it
to the list of clusters most probably associated with GS42-03+37.

  Knowing (or assuming) the distance towards Pup OB3 we can estimate the mass of the gas associated
  with this cluster. Taking all the mass inside the $2.5^{\circ}$ radius from the cluster in the
  whole velocity extent (majority of the mass has $v_{\mathrm{lsr}} \in (25,40)\ \mathrm{kms}^{-1}$,
  see Fig. \ref{pupob3-lblv}) we
  calculate the column density of HI and $\mathrm{H}_2$ using the formulae
 \begin{equation}
  N_{\mathrm{HI}} = A \int T_{\mathrm{HI}} \mathrm{d}v_{\mathrm{HI}} 
  \label{ngas_hi}
 ,\end{equation}
 \begin{equation}
  N_{\mathrm{H_2}} = X \int T_{\mathrm{CO}}\mathrm{d} v_{\mathrm{CO}}
  \label{ngas_h2}
 ,\end{equation} 
 where $T_{\mathrm{HI}}$ is the HI brightness temperature, $T_{\mathrm{CO}}$ is the CO brightness temperature,
 $X$ is the conversion factor for which we use the value
 $X=1.8\times10^{20} cm^{-2}K^{-1}(kms^{-1})^{-1}$, and $A=1.82\times10^{18} cm^{-2}K^{-1}(kms^{-1})^{-1}$.
 Calculated masses are $M_{\mathrm{HI}} = 7.8 \times 10^5 M_{\odot}$ and
 $M_{\mathrm{CO}} = 1.2\times 10^5 M_{\odot}$.
 The total mass of about $9\times 10^5 M_{\odot}$ is close to masses of giant molecular clouds where OB
 associations usually form.
 Therefore the values derived with the assumed distance of 4.5 kpc are not unreasonably high and 
 the assumed distance is not unrealistically large (corresponding masses for the original
 distance of 1.7 kpc would be approximately seven times smaller.

\subsection{When and where were associated clusters created?}

Figure \ref{wall-clusters-names} shows the age gradient along
the wall of the supershell. The oldest clusters are found at low galactic longitudes and
low heliocentric distances, the youngest clusters at high longitudes and large distances.
Two interesting questions arise: Why does such a distribution exist?   And why are
the associated clusters relatively young (much younger than the age of the supershell)?

  Our simulations, as well as other older simulations, for example \citet{tenorio-tagle1987}, show
  that in later stages
of the evolution in the differentially rotating galaxy, the mass in the shell slides along the perimeter
of the `ellipse'. The highest density
is found at the tips, that is, at places near to the major axis. The observed age gradient of clusters
can then be explained by the differences in the growth rates of inhomogeneities:
the medium with higher-density fragments forms stars faster than the medium with low-density fragments.
Older clusters tend to be formed from the medium with the higher column density, which is consistent
with their positions close to the major axis of the supershell.

The second question related to ages of associated clusters is why the onset of star
formation took so long and why about 30 Myr ago, when the supershell was 90 Myr old
(or 50 Myr, if we take the lower limit on the estimated age).
At that time the supershell was not moving supersonically, it was not
sweeping up the new matter and its mass did not grow; in fact, the average column
density decreases with time. The diffential rotation of the Milky Way causes `sliding' of
the material along the walls and accumulation in the `tips', as shown already in
\cite{tenorio-tagle1987} and also observed in our current simulations.
This is relevant to our case: the oldest clusters in our sample were created in the vicinity of the tips.
But this is not the whole story, since clusters are  also found far away from the high-density
tips, in the much-lower-density medium.

One possible explanation of the timescale lies in Fig. \ref{pot-evolution} (right).
In models with the homogenous density or with the thick gaseous disc, the age of 90 Myr
corresponds to the maximum squeezing of the supershell due to the forces acting perpendicularly
to the plane. Correspondingly, the densities in the swept-up wall reached the (local) maxima
and therefore the rate of the gravitational fragmentation was enhanced (or made possible).
There was a similar situation, when the age of the supershell was 50 Myr (i.e. 70 Myr ago),
but we do not find any clusters of the appropriate age, which would appear connected
to the wall.

\subsection{Strange case of NGC 2467-east}

\begin{figure*}
\centering
\includegraphics[angle=-90,width=0.45\linewidth]{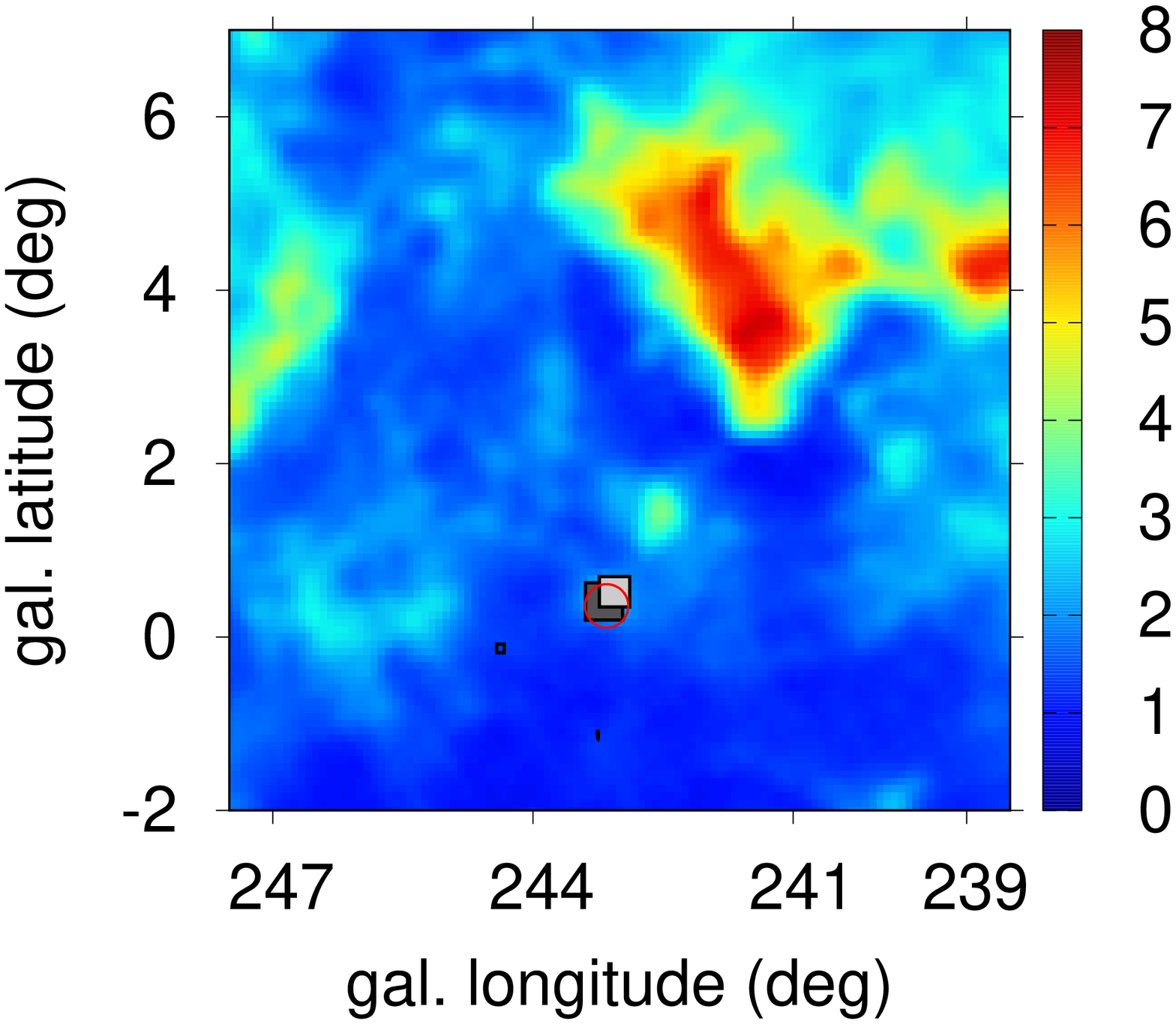}
\includegraphics[angle=-90,width=0.45\linewidth]{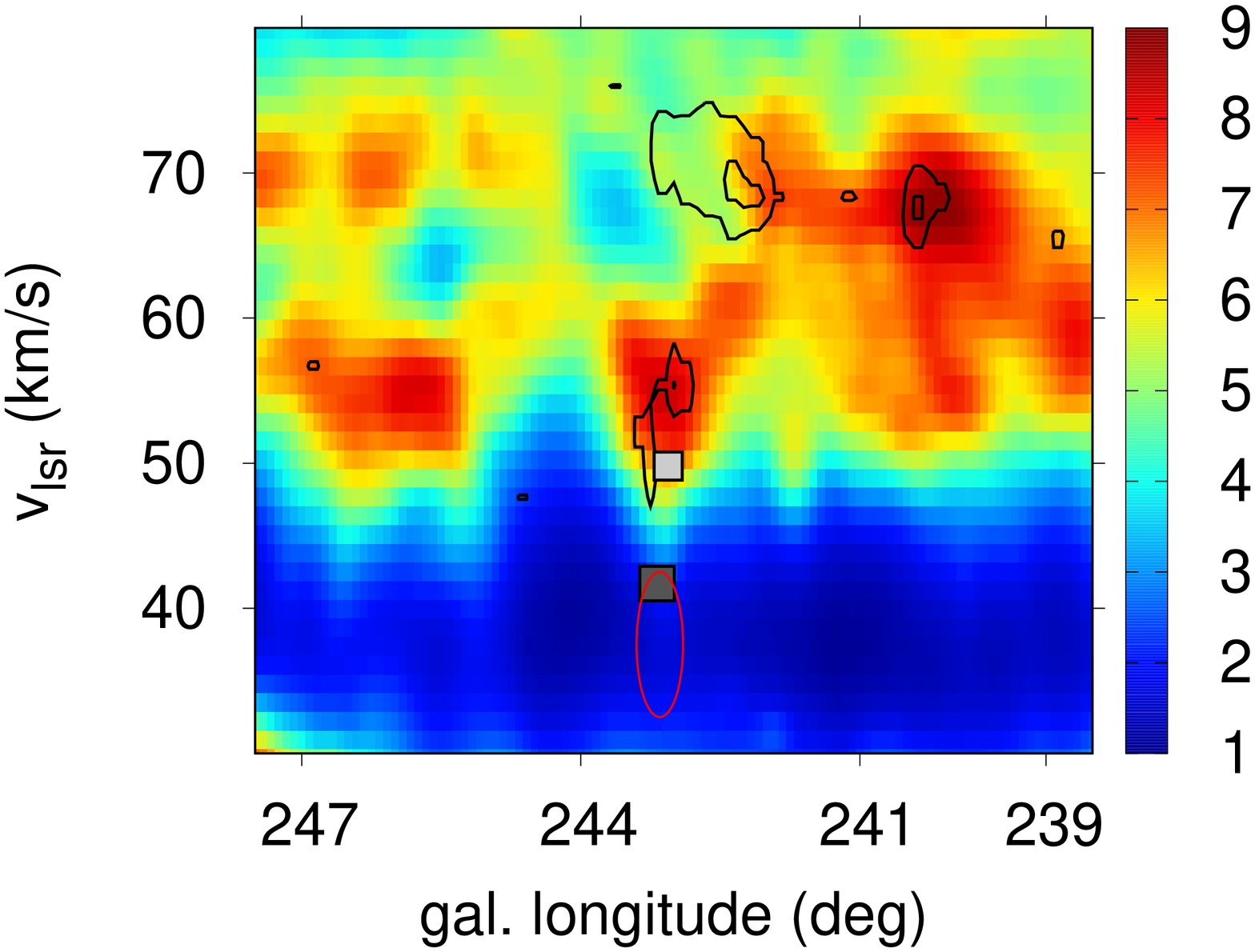}
\caption{NGC 2467-east and the surrounding ISM. The map corresponds to the HI emission,
  contours to the CO emission. The large dark square denotes the position of NGC 2467-east and
  the small light square shows the cluster Haffner 19.
  The red ellipse shows the position of the HII region NGC 2467, its size in the $l$ or $b$ dimension
  is $0.5^{\circ}$, in the $v$ dimension $10 \ kms^{-1}$.
  The left panel is the channel map (velocity interval: $<40.0,44.0>\ kms^{-1}$),
  the right panel is the $lv$-diagram (in the $b$ interval of $<0.2^{\circ},0.6^{\circ}>$).
     }
       \label{ngc2467east-lblv}
\end{figure*}

NGC 2467-east is the young cluster ($\sim$ 6 Myr) with a radial velocity ($v_{\mathrm{lsr}} = 41.7\ kms^{-1}$)
lying at coordinates (l,b) = ($243.18^{\circ},0.41^{\circ}$). It is different from the older NGC 2467 cluster
at very similar coordinates (see \cite{kharchenko+2013}). NGC 2467-east is probably related to
the HII region NGC 2467 (SIMBAD coordinates $243.15^{\circ}$, $0.36^{\circ}$ and the radial velocity
$v_{\mathrm{lsr}} = 37.5\ kms^{-1}$).

The HII region NGC 2467 is an active star-forming region; clusters Haffner 18 and 19 are
believed to lie inside it (https://www.eso.org/public/images/eso0544a/). According to
\cite{kharchenko+2013}, Haffner 19 is a young cluster and according to our analysis it lies
in the distant wall of the supershell and is associated with it. Haffner 18 is, according to the catalogue,
an older cluster with the age of 80 Myr, at a different (nearer) distance than Haffner 19
and not associated with the supershell (though it actually lies close to the near wall).
We reiterate that our `association criteria' are based on distances, not on velocities, since many
clusters do not have measured radial velocities. Therefore, from our point of view, Haffner 19
is (somehow) connected to the HII region NGC 2467, while Haffner 18 is not and lies outside.

NGC 2467-east lies very close to the centre of the supershell GS242-03+37, in an HI- and CO- empty
region. If it is a very young cluster, as given by \cite{kharchenko+2013}, it has nothing to do with
creating the supershell (which is much older), and the real mystery connected with this
cluster is how it came to being in such a low gas density region. Close-ups of $lb$ and $lv$ diagrams
(Fig. \ref{ngc2467east-lblv}) do not help
much. There is some faint emission around the cluster, but really faint. All three objects,
Haffner 19, the HII region NGC 2467, and the cluster NGC 2467-east, have similar $lb$ coordinates;
their radial velocities differ slightly. Haffner 19 lies at the inner edge of the HI wall of the
supershell, HII region and NGC 2467-east lie inwards from the wall; the distance $\Delta r_{shell}$
of the NGC 2467-east is (300-350) pc. To get to its current position, the cluster would have to travel
with a velocity $\geq 50\ kms^{-1}$ (using the age of 6.4 Myr). Such a velocity is relatively high.
It could be lowered slightly if we assume that the star formation did not happen directly in the wall,
but rather in a protruding gaseous `trunk' or a cloud that has wandered inside the supershell.
We know that there is some gas outside the wall (and inside
the supershell), as seen in Fig. \ref{ngc2467east-lblv}, and also known from the presence of the HII
region, but the presence of a relatively large amount of gas needed for the cluster formation is uncertain
(and certainly not observed now).

For an older cluster we could invoke the theory of the star formation triggered by the squeezing of the
preexisting clump by the passing shock front, in times when the supershell GS242-03+37 was younger
and moved supersonically; or perhaps some variant of `pillars of creation' during the early
epoch, when there was hot gas inside the supershell -- it seems there is none there today. For a young
cluster, these scenarios do not work.

Similar cases to NGC 2467-east are clusters Ruprecht 44, FSR 1297, and NGC 2439. Ruprecht 44 and
FSR 1297 do not have measured radial velocities and we only know their distances.
Therefore, especially in the case of Ruprecht 44, we can imagine that a small shift in the
heliocentric distance, comparable to the error in its derivation, could shift the cluster
to the wall (where their formation is much more probable than in the empty hole inside the supershell).
NGC 2439 has a measured radial velocity, which places it inside the hole. It is older than NGC 2467-east
(its age is 18 Myr), but problems with
its formation are similar.

\subsection{An arc of young clusters}

Another interesting feature nicely visible in Fig.\ref{wall-clusters-names} is an arc
of young clusters around the position $l = 233^{\circ}$ and $d_{\mathrm{hc}} = 4\ kpc$.
Clusters FSR 1283 and DBSB 3 (which both lie in the wall of the supershell) and
DBSB 9, FSR 1243, FSR 1228 and NGC 2401 form a beautiful arc, as if blown-out from
the main body of the supershell. DBSB 6 might or might not belong to this arc.
As the supershell is old, the eventual blow-out of gas, from which
the clusters were subsequently formed, could not have happened recently. The theoretical blow-out
could happen due to some very off-centre (supernova) explosion or due to meeting
a preexisting cavity, and this took place during the phase of the supersonic
expansion of the supershell, some 70 - 100 Myr ago.

An alternative explanation is a case of the triggered star formation unrelated to the
GS242-03+37, in a gas expanding from some parent cluster. However, this triggering cluster
is not visible (and it probably should be, because it should not be so old as to move away and get lost), and
we see an age gradient along the arc, compatible with our findings about the age
distribution of clusters along the supershell, but not very compatible with the
simple triggering in the shell.

A chance alignment of clusters is also possible.

\begin{table*}
\caption{Clusters around GS242-03+37 with ages $<$ 120 Myr.}             
\label{table:1}      
\centering
\begin{tabular}{|r l r r r r r r r r|} 
  \hline\hline                        
 ID &  name &  $l$  & $b$ & $d_{\mathrm{hc}}$ & $v_{lsr}$ & age & $\Delta d_{\mathrm{c-w}}$ & Y/N & rem\\
 ~ & ~ & [deg] & [deg] & [kpc] & [km/s] & [Myr] & [kpc] &  & \\
 \hline
  1 & NGC 2345      &   226.58 &    -2.33 &      2.6 &     40.7 &     79.4 &    0.25 &    &    \\
  2 & NGC 2401      &   229.67 &     1.85 &      3.8 &          &     10.0 &    0.21 &    &    \\
  3 & FSR 1228      &   229.88 &    -1.61 &      4.2 &          &      5.6 &    0.40 &    &    \\
  4 & NGC 2414      &   231.40 &     1.93 &      2.9 &     47.5 &     18.6 &    0.03 & y  &    \\
  5 & DBSB 3        &   231.51 &    -4.31 &      3.4 &          &     10.0 &    0.04 & y  &    \\
  6 & FSR 1243      &   231.98 &     1.99 &      4.3 &          &      5.0 &    0.35 &    &    \\
  7 & DBSB 6        &   234.23 &    -0.49 &      4.2 &          &      3.5 &    0.09 &    &    \\
  8 & DBSB 7        &   234.57 &     0.82 &      2.5 &          &     30.9 &    0.15 & y  &    \\
  9 & DBSB 9        &   234.68 &    -0.25 &      4.4 &          &      1.0 &    0.21 &    &    \\
 10 & NGC 2384      &   235.38 &    -2.40 &      2.1 &     30.7 &     13.5 &    0.63 &    &    \\
 11 & NGC 2421      &   236.27 &     0.07 &      2.2 &          &     28.2 &    0.52 &    &    \\
 12 & DBSB 10       &   237.73 &    -0.96 &      2.4 &          &      3.2 &    0.34 &    &    \\
 13 & FSR 1283      &   237.87 &    -3.80 &      4.2 &          &      6.3 &    0.01 & y  &    \\
 14 & Ivanov 6      &   238.48 &    -4.29 &      2.4 &          &      4.0 &    0.27 &    &    \\
 15 & FSR 1297      &   239.57 &    -4.93 &      3.7 &          &      5.0 &    0.29 &    & 1  \\
 16 & Ruprecht 32   &   241.59 &    -0.55 &      4.4 &     64.6 &      5.0 &    0.04 & y  &    \\
 17 & Haffner 16    &   242.09 &     0.47 &      2.7 &     30.4 &     20.0 &    0.11 & y  &    \\
 18 & ESO 429-02    &   242.59 &    -4.16 &      2.1 &          &      7.9 &    0.65 &    &    \\
 19 & Haffner 19    &   243.06 &     0.52 &      4.6 &     49.8 &      4.5 &    0.01 & y  &    \\
 20 & Trumpler 9    &   243.07 &     1.29 &      2.3 &    -34.5 &     63.1 &    0.43 &    &    \\
 21 & Haffner 18    &   243.14 &     0.44 &      2.9 &     41.8 &     79.4 &    0.04 &    &    \\
 22 & NGC 2467-east &   243.18 &     0.41 &      4.0 &     41.7 &      2.6 &    0.35 &    & 1  \\
 23 & NGC 2453      &   243.27 &    -0.94 &      2.4 &     15.5 &     72.4 &    0.35 &    &    \\
 24 & Ruprecht 44   &   245.75 &     0.49 &      4.7 &          &      6.3 &    0.15 &    & 1  \\
 25 & NGC 2439      &   246.45 &    -4.47 &      3.8 &     44.6 &     17.8 &    0.24 &    & 1  \\
 26 & NGC 2489      &   246.72 &    -0.77 &      2.3 &     19.7 &     20.9 &    0.58 &    &    \\
 27 & Haffner 20    &   246.97 &    -0.93 &      2.9 &          &    114.8 &    0.12 &    &    \\
 28 & Haffner 15    &   247.95 &    -4.16 &      2.2 &          &     19.1 &    0.56 &    &    \\
 29 & Bochum 15     &   248.00 &    -5.48 &      2.4 &          &      4.7 &    0.51 &    &    \\
 30 & FSR 1345      &   248.34 &    10.02 &      2.3 &          &      8.9 &    0.51 &    &    \\
 31 & ESO 430-14    &   248.68 &    -0.15 &      2.9 &          &     16.2 &    0.23 &    &    \\
 32 & FSR 1347      &   248.94 &    -4.14 &      2.2 &          &    104.7 &    0.64 &    &    \\
 33 & Haffner 26    &   249.60 &     2.35 &      2.6 &     44.8 &    100.0 &    0.46 &    &    \\
 34 & Ruprecht 55   &   250.72 &     0.81 &      3.6 &     77.6 &      9.3 &    0.04 & y  &    \\
 35 & Ruprecht 59   &   253.07 &     0.93 &      4.0 &     -5.0 &      6.3 &    0.00 & y  &    \\
 36 & ASCC 45       &   253.60 &    -0.34 &      3.0 &     25.4 &     17.8 &    0.37 &    &    \\
 37 & Pup OB3       &   253.93 &    -0.25 &      4.5 &     29.7 &      2.5 &    0.00 & y  &  2 \\
 38 & Pismis 1      &   255.11 &    -0.71 &      5.9 &          &     63.1 &    0.51 &    &    \\
 39 & Ruprecht 154  &   259.58 &    -7.31 &      3.2 &          &     14.1 &    0.58 &    &    \\
 40 & FSR 1397      &   259.90 &     0.34 &      2.1 &          &     60.3 &    1.12 &    &    \\
 41 & DBSB 19       &   259.92 &    -0.05 &      4.3 &          &      1.0 &    0.42 &    &    \\ 
\hline                                        
\end{tabular}
\tablefoot{Columns 1 and 2 show the number and the name of the star cluster.
  Columns 3-7 ($l$, $b$, $d_{\mathrm{hc}}$m $v_{\mathrm{lsr}}$ and `age') give galactic longitudes,
  latitudes, heliocentric distances, radial velocities and ages of clusters. All  values are
  taken from \citet{kharchenko+2013}. Column 8 ($d_{\mathrm{c-w}}$) is the distance of
  the cluster to the wall of the supershell. y in the `Y/N' column 9 indicates if the formation of
  the cluster is very probably connected to the supershell. The last column 10 contains remarks;
  1: the cluster lies inside the supershell. 2: the heliocentric distance of this cluster was
  changed from the catalogue value.
}
\end{table*}

\section{Summary}
\label{sec:summary}

An analysis of the galactic HI supershell GS242-03+37 leads us to the following conclusions.

{Based on the comparison of our numerical model to the HI data we conclude, that
  the supershell GS242-03+37 is an old structure, with the age of 80 Myr or more (our best
  fit is 120 Myr). It could survive so long in the Milky Way because it is luckily placed very
  near the corotation radius of the spiral structure and therefore is not disturbed by the
  passages of the spiral arms, which are otherwise thought of as the main destroyers of larger
  HI shells.}

{The supershell GS242-03+37 is not as energetic a structure as was once thought (and
  may therefore not be a supershell in the strict `energetic' definition). Our estimates give the
  required amount of energy as a few tens of supernovae. While this is still a lot, it is very far from
  previous estimates of hundreds and thousands of supernovae. The discrepancy is given by the effects of the
  differential rotation, which are substantial for such an old structure. Our estimate is
    a lower limit, however, as we do not take into account any leakage of energy into the halo.
}

{There seems to be a correlation between the supershell and the distribution of young
  (< 120 Myr) star clusters: clusters with an age of less than 120 Myr tend to be preferentially
  located in walls of the supershell; no such tendency is found for older clusters.
  The  model of the supershell serves as an example mechanism of how radial velocities (of
    HI data)can be transformed to heliocentric distances (of clusters) and backwards. The age of the model or its
    other properties do not directly influence this comparison.
}

{Clusters, which were most probably created in the wall of the supershell, show an
  age gradient consistent with the densest parts starting fragmentation first. All these
  clusters are younger than 30 Myr. We speculate
  that the onset of star formation is the interplay between the galactic gravitational
  forces --- the accumulation of matter at `tips' due to the galactic shear and the oscillations
  in the $z$-direction --- but we do not propose any detailed model since this would
  require more precise distance and age determinations.}

\begin{acknowledgements}
  We thank the anonymous referee for suggestions and comments, which enriched the original version of the paper.
  This study has been supported by the Albert Einstein Center for Gravitation and Astrophysics
  (the Czech Science Foundation project 14-37086G) and by the project RVO: 67985815.
\end{acknowledgements} 
 
\bibliographystyle{aa} 
\bibliography{gs242_ref}

\end{document}